# Dynamic Modeling, Efficiency Analysis, and Hierarchical Control of Industrial Airborne Pulp Dryers

José M. Campos-Salazar
*Electrical Engineering Department*
*Universidad de Los Lagos*
*Puerto Montt, Chile*

Jorge Gonzalez-Salazar
*Mechanical Engineering Department*
*Universidad de La Frontera*
*Temuco, Chile*

ABSTRACT— ***A comprehensive first-principles dynamic modeling and control framework for industrial airborne pulp dryers is presented. The proposed formulation integrates coupled dry-air transport, water-vapor accumulation, pulp moisture dynamics, psychrometric behavior, convective heat and mass transfer, airflow-pressure interactions, and centrifugal fan electromechanical dynamics within a unified nonlinear state-space representation. An efficiency analysis framework is additionally developed incorporating overall efficiency, thermal efficiency, fan efficiency, and specific moisture evaporation ratio (SMER) as key performance indicators for energetic assessment. To ensure stable operation and enhanced energy utilization, a hierarchical control architecture based on the Direct Synthesis Methodology (DSM) is proposed, combining cascade moisture-temperature and airflow-fan speed control loops, a non-cascade pressure controller, and an efficiency-oriented supervisory optimization layer. The complete nonlinear system was implemented in MATLAB/Simulink and evaluated under multiple thermal, hydraulic, compositional, and mechanical disturbances, including variations in inlet pulp moisture, ambient temperature, production throughput, and industrial vibration effects. Simulation results demonstrate accurate moisture regulation, effective disturbance rejection, stable thermo-fluid operation, and favorable energetic performance. The developed framework achieved an overall efficiency of 85.8%, a thermal efficiency of 86.9%, and an average SMER of 1.39 kg/kWh while maintaining the outlet pulp moisture at the desired operating condition. To the best of the authors' knowledge, this work constitutes one of the first comprehensive dynamic modeling and control studies reported for industrial airborne pulp dryers operating under DryWay-type configurations. The proposed framework provides a proof-of-concept platform suitable for future applications involving digital twins, model predictive control, real-time optimization, and energy-efficient pulp drying systems.***

## 1.0. INTRODUCTION

Drying operations constitute one of the most energy-intensive processes in modern industry and are responsible for a substantial fraction of global industrial thermal energy consumption. It has been estimated that industrial drying systems account for approximately 12–25% of the total energy consumed in manufacturing sectors worldwide, making drying efficiency a critical factor in both economic performance and environmental sustainability [1], [2]. Within the pulp and paper industry, the drying stage represents the largest single consumer of thermal energy, often requiring more than twenty times the energy necessary to remove an equivalent quantity of water through mechanical dewatering methods [3]. Consequently, even modest improvements in dryer operation, heat recovery, airflow management, and process control can translate into significant reductions in fuel consumption, greenhouse gas emissions, and operating costs.

The primary objective of pulp drying is to remove moisture from the wet pulp web while preserving fiber quality, sheet uniformity, brightness, and mechanical integrity. Historically, a variety of drying technologies have been employed in pulp and paper manufacturing, including steam-heated cylinder dryers, flash dryers, rotary dryers, vacuum dryers, through-air dryers, impingement dryers, and hybrid configurations [4]–[8]. Among these alternatives, airborne drying systems have gained increasing industrial relevance due to their capability to provide high drying rates while minimizing mechanical contact with the pulp sheet. In these systems, the pulp web is suspended and transported by high-velocity recirculating air streams, thereby eliminating many of the mechanical constraints associated with conventional contact-based drying technologies [9].

One of the most representative examples of this technology is the DryWay airborne dryer developed for market pulp production. Unlike conventional cylinder-based drying sections, airborne dryers operate through direct convective heat and mass transfer between the pulp sheet and large recirculating air flows. The process typically consists of multiple drying stages where heated air is circulated through enclosed chambers, generating complex interactions among airflow distribution, humidity transport, pressure dynamics, moisture evaporation, thermal storage, and fan performance. Although these systems have been successfully deployed in industrial environments for several decades, their underlying dynamic behavior remains poorly documented in the scientific literature.

Theoretical and experimental investigations of industrial drying systems have traditionally focused on food products, biomass materials, agricultural commodities, wood products, and paper-machine drying sections [5], [10]–[15]. Numerous studies have reported mathematical models describing moisture transport, heat transfer, diffusion mechanisms, and drying kinetics for these applications. For example, the author in [16] developed one of the earliest state-space models for paper-machine drying sections based on coupled mass and energy balances. Similarly, [14] reviewed simulation approaches for multi-cylinder paper dryers and highlighted the necessity of incorporating more physically meaningful descriptions of heat and mass transfer phenomena. More recently, proposed multidimensional models incorporating moisture diffusion through porous paper structures using Richards-type formulations are described by [17], while [18] developed a first-principles dynamic simulator for rotary beet pulp dryers intended for operator training applications.

Despite these significant advances, the majority of published drying models are associated with either steam-heated cylinder dryers or rotary drying systems. Furthermore, most reported models concentrate on local heat-transfer phenomena, moisture diffusion mechanisms, or steady-state performance analysis. Relatively few studies address the complete dynamic interaction among airflow generation, fan dynamics, psychrometric behavior, pressure losses, thermal transport, and moisture evaporation within large-scale industrial drying installations [5], [14], [17], [18].

From a control perspective, drying systems present substantial challenges due to their strongly nonlinear behavior, large transport delays, distributed thermal dynamics, variable operating conditions, and multivariable interactions [19]. Conventional industrial implementations typically employ decentralized PID controllers, cascade structures, and feedforward compensators to regulate temperature, humidity, airflow, and product moisture [20]. While these strategies provide acceptable operational performance, they are often unable to exploit the full energy-saving potential of modern drying facilities.

Consequently, recent research has increasingly explored advanced control methodologies, including adaptive control, robust control, model predictive control, optimization-based supervisory control, and digital-twin-assisted process management [21]–[29], [26], [30]–[33]. For example, authors in [32] demonstrated the application of hierarchical MPC for industrial grain drying systems, achieving significant reductions in energy consumption through the integration of slow supervisory optimization and fast regulatory control layers. Similarly, substantial improvements in drying quality and transient performance through cascade humidity-control architectures are in [33]. These studies collectively demonstrate that advanced control techniques can significantly improve energy utilization when supported by sufficiently accurate process models.

However, the successful implementation of model-based control fundamentally depends on the availability of reliable dynamic models capable of representing the dominant physical mechanisms governing the process [21]. In this regard, the literature reveals a notable gap. Although extensive research exists on paper drying, cylinder drying, tissue drying, flash drying, and rotary drying technologies, no comprehensive dynamic model of industrial airborne pulp dryers has been identified in the open scientific literature. Specifically, no published study has been found that simultaneously incorporates:

- Airflow generation and recirculation dynamics;
- Fan electromechanical behavior;
- Air pressure and duct-flow interactions;
- Convective heat-transfer mechanisms;

- Moisture evaporation dynamics;
- Psychrometric relationships;
- Coupled thermal and moisture balances;
- Hierarchical control structures; and
- Energy-efficiency performance evaluation.

This absence of dynamic modeling frameworks is particularly remarkable considering the widespread industrial utilization of airborne drying systems in modern pulp mills. Moreover, recent reviews of paper drying technologies have explicitly identified the need for improved physical modeling of internal transport phenomena, advanced process simulation capabilities, and model-based optimization methodologies [5].

The increasing industrial interest in digital twins, predictive maintenance, process optimization, and advanced supervisory control further amplifies the importance of developing physically consistent dynamic models for airborne drying systems. Such models constitute essential tools for process understanding, controller development, operational optimization, operator training, fault diagnosis, and future industrial digitalization initiatives.

Motivated by these considerations, this work presents a comprehensive first-principles dynamic model of an industrial airborne pulp drying system. The proposed framework combines mass, energy, pressure, airflow, and moisture balances to describe the coupled dynamics governing dryer operation. In addition, a hierarchical control architecture is developed to illustrate the applicability of the model for future advanced control and optimization studies. The model incorporates psychrometric relationships, convective drying mechanisms, fan dynamics, pressure losses, moisture transport, and energy-efficiency indicators within a unified simulation environment.

It should be emphasized that the primary objective of this study is not experimental validation but rather the development of a physically consistent proof-of-concept modeling framework. Consequently, the presented results focus on dynamic simulations intended to demonstrate model feasibility, analyze process interactions, evaluate energy-efficiency metrics, and establish a foundation for future experimental investigations. To the best of the authors' knowledge, this work represents the first reported dynamic model specifically developed for industrial airborne pulp dryers operating under DryWay-type configurations.

The proposed framework therefore contributes toward closing an important gap between industrial practice and academic research while providing a platform for future developments in digital-twin technology, advanced process control, predictive optimization, and energy-efficient pulp drying operations.

The remainder of this paper is organized as follows. Section II describes the airborne pulp drying process and its principal operating characteristics. Section III develops the nonlinear first-principles dynamic model. Section IV presents the efficiency analysis framework. Section V introduces the hierarchical control architecture and controller design methodology. Section VI discusses the simulation results and performance evaluation. Finally, Section VII summarizes the main conclusions and future research directions.

## 2.0. PROCESS DESCRIPTION

Airborne pulp dryers constitute one of the most critical thermal processing units in modern pulp and paper mills, since they are responsible for reducing the moisture content of the pulp sheet to the required commercial specification prior to cutting, baling, storage, or downstream processing operations [2], [34], [35]. Among current industrial technologies, the DryWay airborne dryer developed by Valmet represents one of the most advanced large-scale drying platforms used in kraft pulp production lines.

According to publicly available industrial information, the DryWay airborne dryer is designed as a multi-level convective drying system capable of processing high production capacities while maintaining elevated energy efficiency, stable web runnability, and optimized airflow distribution throughout the drying structure [34]. The system integrates multiple vertical drying decks, recirculating airflow systems, steam-heated air sections, exhaust networks, and large rotational fan columns distributed along the dryer height [34].

Fig. 1 illustrates a generalized process configuration of an airborne pulp dryer incorporating the principal thermal, fluid dynamic, and mechanical subsystems involved during operation.

The drying process begins when wet pulp sheets leaving the wet end section of the pulp machine enter the dryer with relatively high moisture content. The pulp web is continuously transported through successive drying decks using conveyor and guidance mechanisms specifically designed to maintain sheet stability and minimize flutter phenomena during high-speed operation [6], [35].

Inside the dryer, heated air is injected at high velocity through blow boxes and airflow distribution chambers. The thermal energy transported by the air stream is transferred to the pulp surface primarily by forced convection. As the pulp temperature increases, moisture evaporation occurs within the porous fibrous matrix, generating coupled heat and mass transfer phenomena between the pulp web and the surrounding gas phase [36]–[39].

The evaporated moisture migrates from the internal pore structure of the cellulose sheet toward the surrounding airflow. Consequently, the local humidity ratio of the circulating air increases progressively throughout the drying path. Because of this mechanism, the drying process becomes strongly nonlinear and spatially distributed, since local evaporation rates depend simultaneously on:

- Air temperature,
- Air humidity,
- Airflow velocity,
- Local pressure distribution,
- Pulp consistency,
- Fiber permeability,

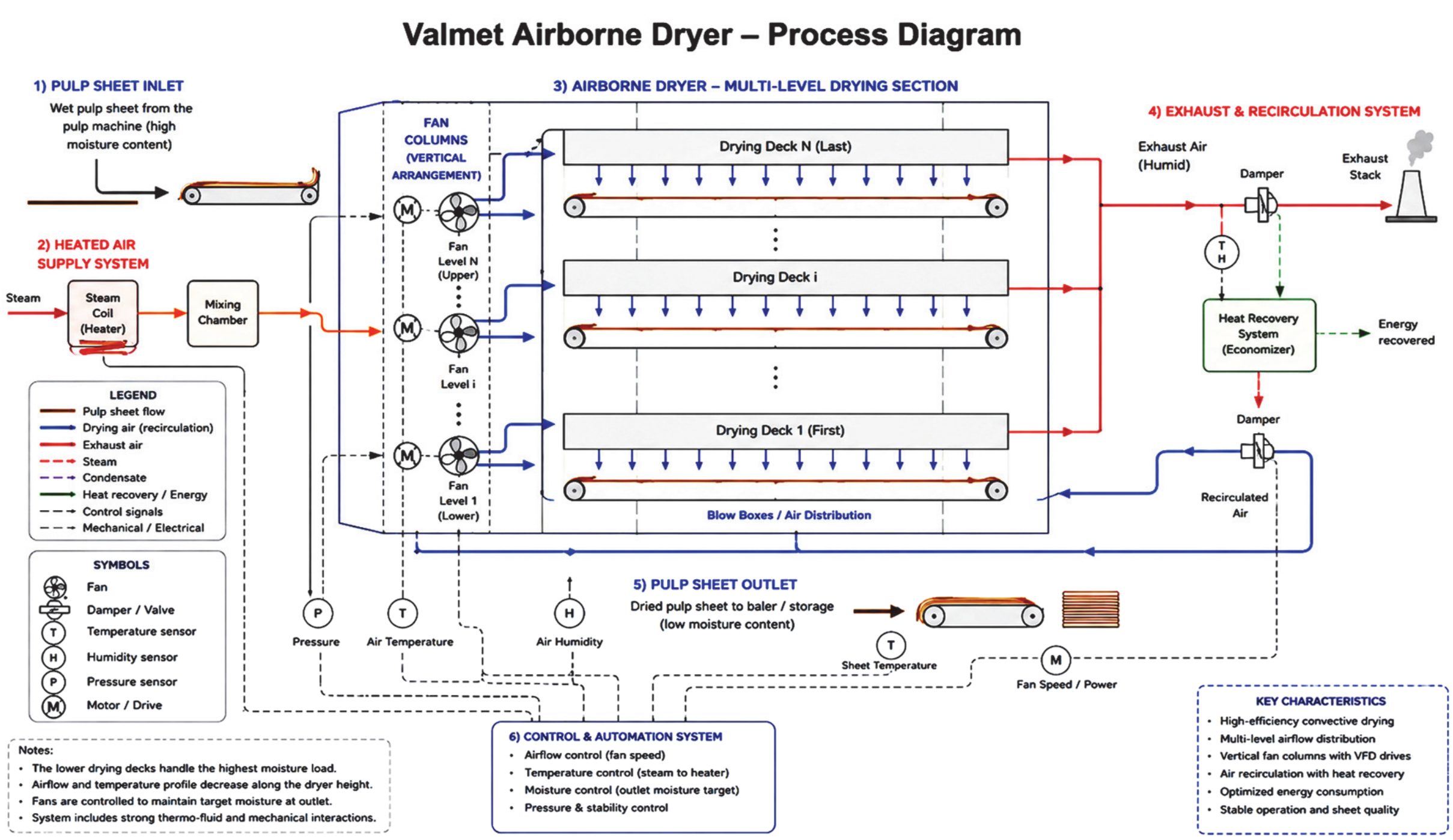

Fig. 1. Generalized process diagram of the Valmet airborne pulp dryer, including the multi-level drying decks, vertical fan columns, airflow recirculation system, heat recovery section, and the principal thermo-fluid and mechanical interactions involved during pulp drying.

- Residence time,
- Heat transfer coefficients,
- Internal moisture diffusion characteristics.

The airborne dryer itself is composed of multiple vertically distributed drying floors or decks connected through recirculating airflow networks. Each deck contains blow boxes, ducting systems, recirculation sections, exhaust outlets, and air distribution plenums. Therefore, the dryer operates as a highly coupled thermo-fluid dynamic network rather than as an isolated drying chamber [1], [4], [10].

One of the most important subsystems of the airborne dryer is the vertical fan column arrangement. These columns contain multiple large-scale centrifugal or axial rotating fans installed along the dryer structure. Each fan level is associated with specific drying decks and is responsible for maintaining the required airflow circulation through the pulp sheet and internal duct systems.

The fans are typically driven by high-power electric motors coupled to converter-frequency drives (VFDs), enabling regulation of rotational speed and airflow generation. From a dynamical perspective, the fan system introduces significant electromechanical interactions into the drying process because the generated airflow depends directly on fan rotational speed, aerodynamic loading conditions, and pressure losses inside the airflow network [40], [41].

The lower drying floors generally operate under the most severe thermodynamic conditions because the entering pulp possesses its highest moisture concentration in these regions. Consequently, the initial drying zones experience:

- Larger evaporation rates,
- Higher latent heat consumption,
- Increased humidity loading,
- Greater airflow requirements,
- Higher pressure losses.

As the pulp travels upward or downstream through the dryer, its moisture concentration decreases progressively. Therefore, the thermal and fluid dynamic requirements of upper drying sections become substantially different from those observed in the initial drying zones [11].

The rotational dynamics of the fan systems are especially important for transient process behavior. Due to the large inertia associated with industrial fan impellers and electric drive systems, changes in motor torque or VFD speed commands do not produce instantaneous airflow variations. Instead, transient acceleration and deceleration dynamics arise, generating additional coupling between airflow generation, pressure distribution, and moisture removal [40].

Furthermore, aerodynamic torque acting on the fan blades depends strongly on airflow resistance throughout the dryer network. Since this resistance changes continuously with pulp moisture content, fouling conditions, airflow recirculation, and production rate, the fan system behaves as a nonlinear electromechanical subsystem coupled directly to the thermodynamic drying process [40].

Another important operational characteristic of airborne pulp dryers is the implementation of recirculating air systems and thermal energy recovery mechanisms. Part of the humid exhaust air may be extracted from the dryer, while another portion is reheated and redirected back into the drying structure. This recirculation mechanism improves energy efficiency but also introduces additional dynamic interactions involving:

- Humidity accumulation,
- Temperature recirculation,

- Pressure balancing,
- Exhaust regulation,
- Heat recovery efficiency.

Consequently, the airborne pulp dryer represents a highly integrated process involving simultaneous interactions among:

- Heat transfer,
- Moisture transport,
- Airflow dynamics,
- Pressure-flow coupling,
- Rotational mechanical systems,
- Psychrometric behavior,
- Distributed transport phenomena,
- Process control systems.

Additionally, the process exhibits significant transport delays because the pulp web requires finite residence time to traverse the drying structure. Disturbances introduced at the dryer inlet may therefore affect outlet moisture conditions only after several minutes. Such transport delays substantially increase the complexity of controller design and process optimization [20], [28].

From an industrial perspective, the principal operational objectives of the airborne dryer include:

- Maintaining uniform outlet moisture content,
- Minimizing thermal energy consumption,
- Reducing electrical power consumption,
- Preventing sheet instability,
- Maximizing production throughput,
- Improving drying efficiency,
- Reducing process variability.

Despite the industrial relevance of airborne pulp drying systems, publicly available first-principles dynamic models incorporating coupled mass balances, energy balances, psychrometric behavior, airflow-pressure interactions, and rotational fan mechanics remain extremely limited in the scientific literature. Most currently available industrial implementations rely primarily on empirical tuning, simplified steady-state calculations, or proprietary internal models developed by equipment manufacturers [5], [34], [40].

Therefore, the development of a comprehensive dynamic modeling framework capable of integrating thermo-fluid dynamics, moisture transport, airflow recirculation, and fan rotational mechanics constitutes a highly significant research opportunity for modern pulp and paper manufacturing systems.

## 3.0. System Modeling

The airborne drying system constitutes a highly coupled thermo-fluid dynamic process in which the transient interactions among the drying gas, suspended pulp fibers, transport ducts, cyclonic separation stages, and turbomachinery determine the overall moisture removal efficiency and process stability. The dynamic behavior of the system is governed by simultaneous conservation of mass, momentum, species concentration, and energy, coupled with pressure-flow interactions and aerodynamic transport phenomena [2], [35], [42]–[44].

In airborne pulp drying systems, wet pulp particles are pneumatically transported through a high-temperature airflow generated by centrifugal fans and heat exchange systems. During transport, convective heat transfer promotes moisture evaporation from the pulp fibers into the gas phase, thereby modifying the thermodynamic properties of both phases simultaneously. Consequently, the system exhibits strongly nonlinear and distributed dynamics due to:

- Variable moisture-dependent pulp properties,
- Airflow compressibility effects,
- Turbulent pressure losses,
- Temperature-dependent evaporation rates,
- Coupled fan-pressure-flow dynamics,
- Nonlinear drying kinetics,
- Distributed thermal transport phenomena.

The complete dynamic representation of the airborne dryer is illustrated through a lumped-parameter multiphysics framework, where the principal state variables are defined as $\mathbf{x}(t) = [M_a(t), M_v(t), M_p(t), T_a(t), T_p(t), \omega_f(t), P_s(t)]^T$, where $M_a(t)$ is the dry air mass inside the dryer in kg, $M_v(t)$ is the water vapor mass in kg, $M_p(t)$ is the dry pulp mass in kg, $T_a(t)$ is the air temperature in K, $T_p(t)$ is the pulp temperature in K, $\omega_f(t)$ is the fan rotational speed in rad/s, and $P_s(t)$ is the internal static pressure in Pa.

The proposed modeling framework combines:

- Thermodynamic conservation equations,
- Convective drying mechanisms,
- Pneumatic transport dynamics,
- Pressure-flow relations,
- Rotational electromechanical fan dynamics.

Such modeling philosophy has been widely employed in industrial drying and transport systems [7], [38], [45]–[47].

### *3.1. Dry Air Mass Balance*

The dry air subsystem provides the fundamental transport mechanism for distributing thermal energy throughout the drying chamber. Its dynamics are governed by the continuity equation

$$\frac{\mathrm{d}M_a(t)}{\mathrm{d}t} = m_a^{in}(t) - m_a^{out}(t) \tag{1}$$

where inflow ($m_a^{in}(t)$) is primarily controlled by fan operation and outflow ($m_a^{out}(t)$) is driven by chamber pressure. Assuming ideal gas behavior [35], the chamber air mass is expressed as

$$M_a(t) = V_d \cdot P_s(t)/(R_a \cdot T_a(t)) \tag{2}$$

Here, $V_d$ and $R_a$ are the effective dryer volume in $m^3$ and the specific gas constant for dry air in J/(kg·K), respectively. Also, (2) reveals nonlinear coupling: higher pressure increases stored mass, while higher temperature reduces it. The outlet flow is modeled as

$$m_a^{out}(t) = C_d \cdot A_d \cdot \sqrt{(2 \cdot \rho_a(t) \cdot (P_s(t) - P_{atm}))} \tag{3}$$

Also from here, $C_d$ is the discharge coefficient, $A_d$ is the equivalent duct area in $m^2$, $\rho_a(t)$ is the air density in $kg/m^3$, and $P_{atm}$ is the atmospheric pressure in Pa.

Equation (3) introduces a square-root dependence on pressure difference. This nonlinearity, typical of compressible duct flows [48], means that small variations in chamber pressure can cause disproportionate changes in outflow, directly influencing drying stability and energy efficiency.

### *3.2. Water Vapor Mass Balance*

The vapor phase accumulates due to moisture evaporation from pulp fibers and transport dynamics. Its balance is given by

$$\frac{\mathrm{d}M_\mathrm{v}(t)}{\mathrm{d}t} = m_\mathrm{v}^{\mathrm{in}}(t) + m_\mathrm{ev}(t) - m_\mathrm{v}^{\mathrm{out}}(t) \tag{4}$$

The variables $m_{ev}(t)$, $m_v^{in}(t)$, and $m_v^{out}(t)$ represent the evaporation rate, the inlet vapor flow rate, and the outlet vapor flow rate in kg/s, respectively.

The evaporation mechanism is modeled through a convective mass transfer relation [36]–[38], [49]:

$$m_{ev}(t) = k_m \cdot A_p \cdot (Y_s(T_p(t)) - Y_a(t)) \tag{5}$$

where $k_m$ is the convective mass transfer coefficient in kg/(m$^2$·s), $A_p$ is the effective pulp surface area in m$^2$, and the humidity variables: $Y_s(T_p(t))$ and $Y_a(t)$ corresponding to the bulk air humidity ratio and the saturation humidity ratio at pulp temperature, respectively.

The humidity ratio is defined as:

$$Y_a(t) = M_v(t)/M_a(t) \tag{6}$$

while the saturation humidity ratio may be approximated using psychrometric relations [36]–[38]:

$$Y_s(T_p(t)) = 0.622 \cdot P_{sat}(T_p(t))/(P_s(t) - P_{sat}(T_p(t))) \tag{7}$$

where $P_{sat}(T_p(t))$ is the saturation vapor pressure.

This formulation couples thermodynamics with drying kinetics, highlighting how pulp temperature directly influences evaporation rates. As pulp warms, saturation vapor pressure rises, increasing the driving force for evaporation, a phenomenon extensively documented in convective drying studies [2], [44], [50].

### *3.3. Dry Pulp Mass Balance*

The dry pulp mass transport dynamics are represented by:

$$\frac{\mathrm{d}M_\mathrm{p}(t)}{\mathrm{d}t} = m_\mathrm{p}^{\mathrm{in}}(t) - m_\mathrm{p}^{\mathrm{out}}(t) \tag{8}$$

$m_p^{in}(t)$ and $m_p^{out}(t)$ are the inlet dry pulp flow rate and the outlet dry pulp flow rate in kg/s.

The moisture content of the pulp is defined as:

$$X_p(t) = M_w(t)/M_p(t) \tag{9}$$

where $M_w(t)$ represents liquid water retained in the fibers.

The liquid moisture variation satisfies:

$$\frac{\mathrm{d}M_\mathrm{w}(t)}{\mathrm{d}t} = m_\mathrm{w}^{\mathrm{in}}(t) - m_\mathrm{ev}(t) - m_\mathrm{w}^{\mathrm{out}}(t) \tag{10}$$

which directly couples evaporation with fiber moisture. This nonlinear relation reflects the fact that drying slows as fibers lose water and surface area for evaporation decreases. Such behavior is consistent with industrial drying kinetics, where moisture removal rates decline as fibers approach equilibrium moisture content [2], [5].

### *3.4. Air Energy Balance*

The thermal dynamics of the drying gas follow the first law of thermodynamics [35]–[38], [49]:

$$c_p^a \cdot \frac{d\left(M_a(t) \cdot T_a(t)\right)}{dt} = q_{in}(t) - q_{ev}(t) - q_{loss}(t) + h_{in}(t) - h_{out}(t) \tag{11}$$

From here, $c_p{}^a$ is the specific heat of air at constant pressure, $q_{in}(t)$, $q_{ev}(t)$, and $q_{loss}(t)$ are the supplied thermal power, the latent heat consumed by evaporation, and the thermal losses, respectively. Finally, $h_{in}(t)$ and $h_{out}(t)$ are the enthalpy transport rates.

The latent heat consumption is:

$$q_{ev}(t) = \lambda_v \cdot m_{ev}(t) \tag{12}$$

where $\lambda_v$ denotes latent heat of vaporization.

Thermal losses through dryer walls may be modeled as:

$$q_{loss}(t) = U_d \cdot A_d \cdot (T_a(t) - T_{amb}) \tag{13}$$

$U_d$ and $A_d$ denote the overall heat transfer coefficient and the dryer wall area, respectively. $T_{amb}$ is the ambient temperature.

This balance shows that heat input must simultaneously sustain evaporation and compensate for losses, making insulation and heat supply critical to efficiency. In practice, inadequate insulation leads to significant energy penalties, while excessive heat input risks pulp degradation [7], [36]–[38], [49].

### *3.5. Pulp Thermal Dynamics*

The pulp thermal behavior is represented by:

$$c_p^p \cdot \frac{d\left(M_p(t) \cdot T_p(t)\right)}{dt} = q_{conv}(t) - q_{ev}(t) \tag{14}$$

In this case, $c_p{}^p$ is the pulp specific heat and $q_{conv}(t)$ is the convective heat transfer.

The convective heat exchange between air and fibers is:

$$q_{conv}(t) = h_c \cdot A_p \cdot (T_a(t) - T_p(t)) \tag{15}$$

where $h_c$ is the convective heat transfer coefficient.

Equation (15) establishes strong coupling: air cools as pulp heats, and turbulence (higher $h_c$) accelerates drying but increases fan energy demand. The trade-off between drying rate and energy consumption is a central design consideration in convective drying systems [2], [49], [51].

### *3.6. Airflow Pressure Dynamics*

The pressure evolution inside the dryer results from compressible airflow transport and fan operation:

$$\frac{dP_s(t)}{dt} = \frac{R_a}{V_d} \cdot T_a(t) \cdot \left(m_a^{in}(t) - m_a^{out}(t)\right) \tag{16}$$

This equation reveals that pressure dynamics are directly influenced by:

- Air density,
- Temperature,
- Fan-induced flow,
- Hydraulic resistance of ducts and cyclones.

The distributed pressure losses are modeled through Darcy-Weisbach-type relations [41]:

$$\Delta P_d(t) = 0.5 \cdot (f_d \cdot L_d / D_d) \cdot \rho_a(t) \cdot v_d^2(t) \tag{17}$$

From here, $f_d$, $L_d$, and $D_d$ are friction factor, the duct length, and the hydraulic diameter, respectively. Also, $v_d(t)$ is the airflow velocity.

These quadratic losses highlight sensitivity to duct geometry and airflow velocity. Fan pressure rise must balance them to maintain stable operation, and improper duct design can lead to oscillatory pressure dynamics and unstable drying [41].

### *3.7. Centrifugal Fan Dynamics*

The centrifugal fan constitutes the principal driving mechanism of the pneumatic drying process. Its rotational dynamics are represented by:

$$\frac{\mathrm{d}\omega_{\mathrm{f}}(t)}{\mathrm{d}t} = \frac{1}{J_{\mathrm{f}}}\cdot\left(\tau_{\mathrm{m}}(t) - \tau_{\mathrm{a}}(t) - B_{\mathrm{f}}\cdot\omega_{\mathrm{f}}(t)\right) \tag{18}$$

where $J_{\mathrm{f}}$ and $B_{\mathrm{f}}$ are the rotational inertia and the viscous friction coefficient, respectively. Moreover, $\tau_{\mathrm{m}}(t)$ and $\tau_{\mathrm{a}}(t)$ are the motor and the aerodynamic load torque, respectively.

The aerodynamic load torque is proportional to airflow demand:

$$\tau_{\mathrm{a}}(t) = K_{\mathrm{a}}\cdot\omega_{\mathrm{f}}^{2}(t) \tag{19}$$

while the generated airflow satisfies the fan affinity laws [41]:

$$m_{\mathrm{a}}^{\mathrm{in}}(t) = K_{\mathrm{f}}\cdot\omega_{\mathrm{f}}(t)\cdot\sqrt{(P_{\mathrm{f}}(t)/\rho_{\mathrm{a}}(t))} \tag{20}$$

From here, $K_{\mathrm{f}}$ and $P_{\mathrm{f}}(t)$ are the fan flow coefficient and the developed fan pressure, respectively. The fan pressure characteristic can be approximated by:

$$P_{\mathrm{f}}(t) = P_{\mathrm{f0}} - K_{\mathrm{p}}\cdot {m_{\mathrm{a}}^{\mathrm{in}}}^{2}(t) \tag{21}$$

These relations capture the nonlinear quadratic pressure–flow curve typical of centrifugal fans, where motor torque must continuously overcome aerodynamic load to sustain desired airflow. The electromechanical coupling between motor torque and aerodynamic resistance is a critical factor in fan control strategies [41].

## 4.0. EFFICIENCY EXPRESSION

The efficiency of the airborne pulp dryer constitutes one of the most important performance indicators since it quantifies the capability of the system to convert the thermal and electrical energy supplied to the process into useful moisture removal from the pulp fibers. Unlike conventional drying systems, the airborne dryer exhibits simultaneous interactions among heat transfer, mass transfer, airflow generation, pressure transport, moisture evaporation, and fan electromechanical dynamics. Consequently, an efficiency formulation must account for both thermal and mechanical energy contributions as well as the latent energy associated with moisture removal [5], [7], [36]–[38], [44], [51].

From a thermodynamic perspective, the useful output of the drying process corresponds to the latent heat required to evaporate moisture from the pulp web, whereas the energy inputs are associated with:

- Steam heating systems,
- Air-heating coils,
- Fan motor power consumption,
- Auxiliary equipment,
- Air recirculation systems.

Therefore, the overall efficiency of the airborne dryer may be defined as

$$\eta_{\mathrm{tot}}(t) = q_{\mathrm{useful}}(t)/(q_{\mathrm{in}}(t) + P_{\mathrm{f}}(t)) \tag{22}$$

where: $\eta_{\mathrm{tot}}(t)$ is the overall dryer efficiency, $q_{\mathrm{useful}}(t)$ is the useful evaporation energy, $q_{\mathrm{in}}(t)$ is the thermal power supplied by the heating system, and $P_{\mathrm{f}}(t)$ is the electrical power consumed by the fan system.

Equation (22) establishes a global energetic performance indicator incorporating both thermal and mechanical energy consumption. Unlike conventional dryer efficiencies that neglect airflow generation costs, the proposed formulation explicitly includes fan power requirements, which may represent a significant fraction of the total operating cost in airborne drying systems.

### *4.1. Useful Drying Energy*

The primary objective of the airborne pulp dryer is the removal of moisture from the pulp web through evaporation. Consequently, the useful energy associated with the drying process corresponds to the latent heat required to transform liquid water retained within the cellulose fiber structure into water vapor that is subsequently transported by the drying airflow. From a thermodynamic perspective, this latent energy represents the desired energetic output of the process and constitutes the principal mechanism through which the moisture content of the pulp is reduced [2], [35]–[37], [52].

The useful drying energy may therefore be expressed as

$$q_{useful}(t) = \lambda_v \cdot m_{ev}(t) \quad (23)$$

where $\lambda_v$ is the latent heat of vaporization and $m_{ev}(t)$ is the moisture evaporation rate. Equation (23) establishes a direct relationship between moisture removal and useful energy utilization, indicating that the energetic effectiveness of the drying process increases proportionally with the evaporation rate. Since evaporation constitutes the primary purpose of the dryer, $q_{useful}(t)$ may be interpreted as the portion of the supplied energy that directly contributes to product drying.

Substituting (23) into (22) yields

$$\eta_{tot}(t) = \lambda_v \cdot m_{ev}(t)/(q_{in}(t) + P_f(t)) \quad (24)$$

where $q_{in}(t)$ denotes the thermal power supplied by the heating system and $P_f(t)$ represents the electrical power consumed by the airflow generation subsystem. Equation (24) demonstrates that the overall efficiency of the airborne dryer depends simultaneously on the evaporation rate, the thermal energy supplied to the process, and the electrical energy required to maintain airflow circulation. Consequently, maximum efficiency is achieved when elevated moisture removal rates are obtained with minimum thermal and mechanical energy consumption.

Moreover, (24) reveals that increases in evaporation rate do not necessarily imply improved efficiency if they are accompanied by disproportionately large increases in heating requirements or fan power consumption. This observation is particularly important in airborne dryers because excessive airflow rates may increase electrical consumption while providing only marginal improvements in moisture removal once the psychrometric driving force begins to decrease. Therefore, optimization of airflow distribution, fan operation, heat-transfer mechanisms, and moisture transport dynamics becomes essential for maximizing energetic performance [2], [5], [40], [41], [52].

### *4.2. Thermal Efficiency*

The thermal efficiency evaluates the capability of the heating subsystem to convert supplied thermal energy into useful latent heat associated with moisture evaporation. This performance indicator isolates the thermal behavior of the drying process from the electrical power consumed by the fan network and provides a direct measure of the effectiveness of the heat-transfer mechanisms occurring within the dryer [2], [5], [6], [35]–[37], [40], [41], [49], [51]–[54].

The thermal efficiency may be defined as

$$\eta_{th}(t) = \lambda_v \cdot m_{ev}(t)/q_{in}(t) \quad (25)$$

which represents the ratio between the useful latent heat utilized for moisture evaporation and the total thermal energy supplied to the process. In an ideal drying system, all supplied thermal energy would be converted into latent heat. However, practical industrial dryers always exhibit efficiencies below unity because part of the supplied energy is dissipated through unavoidable thermal losses and irreversible heat-transfer mechanisms.

The principal energy dissipation mechanisms include heat losses through the dryer enclosure, sensible heating of the drying air, sensible heating of the pulp fibers, thermal energy transported away by the exhaust stream, and transient thermal accumulation within the process. Consequently, the thermal efficiency can be interpreted as a direct indicator of the capability of the dryer to convert supplied heat into productive moisture removal rather than parasitic energy consumption.

The heat losses through the dryer enclosure may be approximated as

$$q_{loss}(t) = U_d \cdot A_d \cdot (T_a(t) - T_{amb}) \quad (26)$$

where $U_d$ is the overall heat-transfer coefficient, $A_d$ is the effective heat-transfer area, and $T_{amb}$ is the ambient temperature. Equation (26) indicates that thermal losses increase proportionally with the temperature difference between the drying chamber and the surrounding environment, emphasizing the importance of adequate thermal insulation in industrial airborne dryers operating at elevated temperatures.

By incorporating thermal losses and sensible heating requirements, (25) may be reformulated as

$$\eta_{th}(t) = 1 - (q_{loss}(t) + q_{sens}(t))/q_{in}(t) \quad (27)$$

where $q_{sens}(t)$ represents the sensible energy required to increase the temperature of both the drying air and the pulp fibers. Equation (27) clearly demonstrates that thermal efficiency is fundamentally governed by the competition between useful latent heat utilization and parasitic thermal losses. Consequently, any reduction in wall heat losses, exhaust enthalpy losses, or excessive sensible heating directly contributes to improving the thermal performance of the dryer.

Furthermore, (27) provides a rigorous thermodynamic justification for the implementation of thermal insulation systems, exhaust heat-recovery units, air-recirculation strategies, and optimized airflow management. Such technologies reduce the magnitude of $q_{loss}(t)$ while increasing the fraction of supplied energy that contributes directly to moisture evaporation. This principle forms the basis of modern high-efficiency airborne drying systems employed throughout the pulp and paper industry and constitutes one of the principal design criteria used in contemporary industrial drying technologies [2], [34], [36]–[38], [51], [53].

### *4.3. Fan Efficiency*

The airflow generation subsystem plays a fundamental role in airborne drying systems because it is responsible for transporting thermal energy, removing evaporated moisture, maintaining the required pressure gradients, and promoting convective heat and mass transfer between the drying air and the pulp web. Consequently, the performance of the fan system directly influences both the drying capacity and the overall energetic efficiency of the process [40], [41].

The mechanical power delivered by the fan shaft may be expressed as

$$P_{mech}(t) = \tau_m(t) \cdot \omega_f(t) \quad (28)$$

$\tau_m(t)$ and $\omega_f(t)$ denote the motor torque and the fan angular velocity, respectively. Equation (28) represents the instantaneous mechanical energy transferred from the motor to the rotating fan assembly and constitutes the primary source of airflow generation within the dryer.

The aerodynamic power effectively transferred to the airflow is given by

$$P_{air}(t) = \Delta P_d(t) \cdot q_a(t) \quad (29)$$

where $\Delta P_d(t)$ is the pressure rise produced by the fan and $q_a(t)$ is the volumetric airflow rate. This expression quantifies the useful mechanical energy converted into airflow transport through the drying system.

Consequently, the fan efficiency may be defined as

$$\eta_f(t) = P_{air}(t)/P_{mech}(t) \quad (30)$$

which measures the effectiveness of converting shaft power into useful airflow generation. High values of $\eta_f(t)$ indicate efficient operation near the fan best-efficiency point, whereas lower values suggest excessive hydraulic losses, flow recirculation, flow separation, or operation away from the design conditions.

Since the airflow directly influences the heat-transfer coefficient, mass-transfer coefficient, pressure distribution, and moisture transport mechanisms developed in (5), (15), and (17), fan efficiency becomes strongly coupled with the overall drying performance. Therefore, optimization of fan rotational speed, duct geometry, and airflow distribution is essential for minimizing electrical consumption while maintaining adequate drying capacity [40], [41].away from the fan best-efficiency point.

## 5.0. CONTROL SYSTEM ANALYSIS

The airborne drying process constitutes a highly coupled thermo-fluid dynamic system characterized by strong interactions among moisture transport, heat transfer, airflow distribution, pressure regulation, and energy consumption. Variations in pulp feed moisture, ambient conditions, airflow characteristics, and thermal input directly affect the evaporation rate and consequently the final pulp moisture content. Furthermore, the presence of multiple manipulated variables and competing operational objectives introduces a multivariable control problem in which product quality and energy efficiency must be simultaneously optimized [2], [5], [28], [41], [54]–[56].

To ensure stable operation and maximize drying performance, a hierarchical control architecture is proposed, combining cascade and non-cascade control loops designed using the direct synthesis methodology (DSM). The proposed strategy employs conventional regulatory control at the lower level and efficiency-based optimization at the supervisory level. The lower control layer guarantees moisture, temperature, airflow, and pressure regulation, whereas the upper layer continuously adjusts operating conditions to maximize the overall dryer efficiency. The complete proposed control architecture is illustrated in Fig. 2.

Fig. 2 illustrates the proposed hierarchical control architecture developed for the airborne pulp dryer. The overall structure combines cascade and non-cascade control loops designed using the DSM, together with a supervisory optimization layer responsible for maximizing the energetic performance of the drying process. The architecture was developed to simultaneously satisfy product-quality requirements and energy-efficiency objectives while maintaining stable operation under process disturbances.

At the highest level, the supervisory optimization layer continuously evaluates the principal performance indicators of the dryer, namely $\eta_{\text{tot}}(t)$, and $\eta_{\text{th}}(t)$. These performance metrics are calculated from the dynamic process measurements and incorporated into the economic objective function defined in (56). Based on the optimization results, the supervisory controller generates the optimal drying-air temperature reference $T_a^*(t)$ and airflow reference $q_a^*(t)$, which are subsequently transmitted to the lower regulatory control layer. In this manner, the supervisory controller continuously adjusts the operating point of the dryer to maximize moisture removal while minimizing thermal and electrical energy consumption.

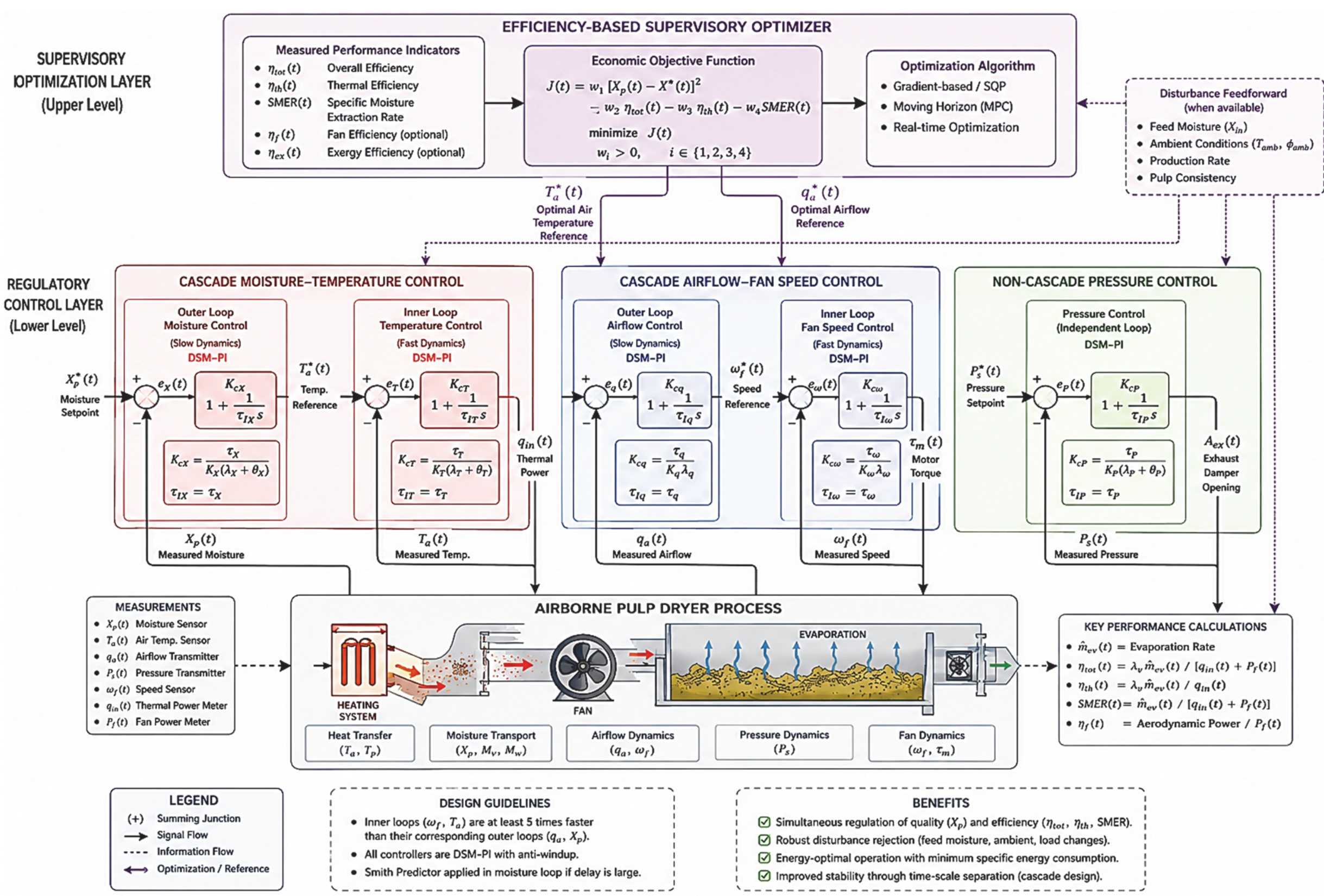

Fig. 2. Proposed hierarchical control architecture for the airborne pulp dryer, combining DSM-based cascade moisture–temperature and airflow–fan speed control loops, a non-cascade pressure controller, and a supervisory efficiency optimizer that maximizes drying performance while minimizing energy consumption.

The regulatory layer consists of three principal control subsystems. The first subsystem corresponds to the cascade moisture–temperature control structure. Since moisture removal is strongly dependent on drying-air temperature, the outer moisture controller regulates the final pulp moisture content $X_p(t)$ by generating the reference signal for the inner temperature-control loop. The inner temperature controller subsequently manipulates the thermal power supplied to the heating system, $q_{in}(t)$, ensuring rapid rejection of thermal disturbances and improved moisture regulation. This cascade arrangement significantly improves control performance because the temperature dynamics are considerably faster than the moisture dynamics.

The second subsystem corresponds to the cascade airflow–fan-speed control structure. In this configuration, the outer airflow controller regulates $q_a(t)$ by generating the reference rotational speed for the fan system. The inner fan-speed controller then manipulates the motor torque $\tau_m(t)$ to achieve the desired rotational speed $\omega_f(t)$. Because the fan rotational dynamics are substantially faster than the airflow dynamics, the cascade structure improves disturbance rejection and provides more accurate airflow regulation throughout the dryer.

The third subsystem consists of an independent non-cascade pressure-control loop. Chamber pressure $P_s(t)$ is regulated through the exhaust damper opening $A_{ex}(t)$, allowing the dryer to maintain the desired internal pressure conditions independently of the airflow and moisture-control subsystems. This configuration minimizes undesirable interactions between pressure regulation and the remaining control loops while ensuring stable airflow distribution throughout the drying chambers.

The controlled process block shown in Fig. 2 represents the complete airborne pulp dryer model developed in Sections 4 and 5. This block contains the nonlinear interactions among heat transfer, moisture transport, airflow generation, pressure dynamics, psychrometric behavior, and fan rotational dynamics. The principal state variables include $T_a(t)$, $T_p(t)$, $X_p(t)$, $P_s(t)$, $\omega_f(t)$, $M_v(t)$, and $M_w(t)$. These variables evolve according to the coupled mass, energy, and momentum balances previously derived.

Finally, the feedback signals measured throughout the dryer provide real-time information to the corresponding controllers, thereby closing the regulatory loops. The resulting hierarchical architecture integrates product-quality control, airflow management, pressure stabilization, and efficiency optimization within a unified framework. Consequently, the proposed control strategy enables simultaneous regulation of outlet pulp moisture, drying-air temperature, chamber pressure, and airflow distribution while continuously maximizing the overall energetic performance of the airborne drying process. This combination of regulatory and supervisory control provides a suitable foundation for future implementation of advanced optimization and model-predictive control strategies in industrial airborne pulp dryers [6], [35]–[38], [49], [49], [51], [53], [54].

### *5.1. Control Objectives*

The primary controlled-variable vector of the airborne dryer is defined as $\mathbf{y}(t) = [X_p(t), T_a(t), P_s(t), q_a(t)]^T$, where $X_p(t)$, $T_a(t)$, $P_s(t)$, $q_a(t)$ are the pulp moisture content, the drying air temperature, the chamber pressure, and the drying air flowrate, respectively. The corresponding manipulated-variable vector are $\mathbf{u}(t) = [q_{in}(t), \tau_m(t), A_{ex}(t)]^T$, where $q_{in}(t)$, $\tau_m(t)$, $A_{ex}(t)$ are the thermal energy supplied by the heating system, the fan motor torque, and the exhaust damper opening are, respectively.

The control system must satisfy the following operational objectives:

1. Maintain outlet pulp moisture at its desired value.
2. Regulate drying air temperature.
3. Maintain stable chamber pressure.
4. Ensure adequate airflow distribution.
5. Maximize thermal and overall drying efficiencies.
6. Minimize specific energy consumption.

7. Reject disturbances caused by feed moisture variations.

*5.2. Direct Synthesis Controller Design*

The DSM was selected due to its simplicity, robustness, and ability to directly impose desired closed-loop dynamics [19], [28], [55]–[57].

For a generic process represented by

$$G_p(s) = K_p \cdot e^{-\theta s}/(\tau_p \cdot s + 1) \tag{31}$$

the desired closed-loop transfer function is defined as

$$G_{cl}(s) = e^{-\theta s}/(\lambda \cdot s + 1) \tag{32}$$

where $\lambda$ represents the desired closed-loop time constant. The DSM controller is obtained from

$$G_c(s) = G_{cl}(s)/(G_p(s) \cdot [1 - G_{cl}(s)]) \tag{33}$$

which yields the well-known PI form

$$G_c(s) = K_c \cdot (1 + 1/(\tau_I \cdot s)) \tag{34}$$

with

$$K_c = \tau_p/(K_p \cdot [\lambda + \theta]) \tag{35}$$

and

$$\tau_I = \tau_p \tag{36}$$

The tuning parameter $\lambda$ determines the trade-off between response speed and robustness.

*5.1. Cascade Airflow Control System*

The airflow generation subsystem exhibits two significantly different time scales:

1. Fan rotational dynamics.
2. Airflow dynamics.

Consequently, a cascade control structure is adopted.

a) Inner Loop: Fan Speed Control

The fan rotational dynamics are described by (37). Linearization around an operating point yields the expression in (38).

$$\frac{d\omega_f(t)}{dt} = \frac{1}{J_f} \cdot \left(\tau_m(t) - B_f \cdot \omega_f(t) - K_a \cdot \omega_f^2(t)\right) \tag{37}$$

$$G_\omega^\tau(s) = \Delta\omega_f(t)/\Delta\tau_m(t) = K_\omega/(\tau_\omega \cdot s + 1) \tag{38}$$

The corresponding DSM controller is

$$G_c^\omega(s) = K_c^\omega/(1 + 1/(\tau_I^\omega \cdot s)) \tag{39}$$

with

$$K_c^\omega = \tau_\omega/(K_\omega \cdot \lambda_\omega) \tag{40}$$

and

$$\tau_I^\omega = \tau_\omega \tag{41}$$

b) Outer Loop: Airflow Control

The airflow dynamics are approximated as

$$G_q(s) = K_q/(\tau_q \cdot s + 1) \tag{42}$$

The outer airflow controller is therefore

$$G_c^q(s) = K_c^q/(1 + 1/(\tau_I^q \cdot s)) \tag{43}$$

where

$$K_c^q = \tau_q/(K_q \cdot \lambda_q) \tag{44}$$

and

$$\tau_I^q = \tau_q \tag{45}$$

The inner speed loop is designed to be at least five times faster than the outer airflow loop.

*5.2. Cascade Moisture-Temperature Control System*

Moisture control constitutes the most critical quality variable of the drying process. Since moisture removal is strongly affected by air temperature, a cascade structure is employed.

a) Inner Loop: Air Temperature Control

The thermal dynamics are represented by

$$G_T(s) = K_T \cdot e^{-\theta_T \cdot s}/(\tau_T \cdot s + 1) \tag{46}$$

The DSM controller becomes

$$K_c^T = \tau_T/(K_T \cdot [\lambda_T + \theta_T]) \tag{47}$$

and

$$\tau_I^T = \tau_T \tag{48}$$

b) Outer Loop: Moisture Control

The moisture dynamics can be approximated by

$$G_X(s) = K_X \cdot e^{-\theta_X \cdot s}/(\tau_X \cdot s + 1) \tag{49}$$

The moisture controller is

$$K_c^X = \tau_X/(K_X \cdot [\lambda_X + \theta_X]) \tag{50}$$

and

$$\tau_I^X = \tau_X \tag{51}$$

Due to the large transport delay associated with moisture measurement, a Smith Predictor may be incorporated to improve performance [58].

c) Non-Cascade Pressure Control System

Pressure regulation is treated independently from the moisture and thermal control loops. The chamber pressure dynamics can be represented as

$$G_P(s) = K_P/(\tau_P \cdot s + 1) \tag{52}$$

The pressure controller is obtained through DSM as

$$K_c^P = \tau_P/(K_P\cdot[\lambda_P + \theta_P]) \tag{53}$$

and

$$\tau_I^P = \tau_P \tag{54}$$

Pressure control is implemented through the exhaust damper opening $A_{ex}$.

d) Efficiency-Based Supervisory Control

Although moisture control guarantees product quality, it does not necessarily maximize energy utilization. Therefore, a supervisory optimization layer is proposed based on the efficiency indicators derived previously. The overall drying efficiency is defined in (24) and the thermal efficiency is defined in (25). Moreover, the specific moisture extraction rate (SMER) can be defined as

$$SMER(t) = m_{ev}(t)/(q_{in}(t) + P_f(t)) \tag{55}$$

The supervisory controller continuously evaluates these performance indicators and computes optimal temperature and airflow references. An economic objective function is defined as

$$J(t) = w_1\cdot(X_p(t) - X^*(t))^2 - w_2\cdot\eta_{tot}(t) - w_3\cdot\eta_{th}(t) - w_4\cdot SMER(t) \tag{56}$$

where $w_i$ are weighting coefficients. From here, $i \in \{1, 2, 3, 4\}$. Minimization of (56) simultaneously:

- minimizes moisture tracking error,
- maximizes evaporation rate,
- minimizes energy consumption,
- maximizes overall dryer efficiency.

## 6.0. Simulation Results

The proposed airborne pulp dryer model was implemented in MATLAB using the nonlinear thermo-fluid dynamic framework developed in Sections 4–6. Moreover, Table 1 list the main simulation parameters. The simulation environment integrates moisture transport, heat transfer, psychrometric behavior, airflow generation, chamber-pressure dynamics, fan rotational mechanics, DSM-based cascade and non-cascade control loops, and the efficiency-oriented supervisory optimization layer. The complete simulation was performed over a horizon of 3600 s and required only 3.94 s of computational time, demonstrating the numerical efficiency of the proposed formulation and its suitability for future real-time optimization, digital-twin implementation, and advanced supervisory control applications.

The quantitative indicators summarized in Tables 2 and 3 demonstrate that the proposed hierarchical control architecture successfully regulates the most important process variables while simultaneously maintaining favorable energetic performance. The outlet pulp moisture converges to 0.1201 kg/kg dry pulp, indicating that the quality objective is achieved. Similarly, the final drying-air temperature reaches 409.83 K, while the psychrometric driving force converges to 0.2039 kg/kg, providing favorable conditions for sustained moisture removal. From an energetic standpoint, the overall efficiency reaches 85.80%, the thermal efficiency converges to 86.96%, and the SMER attains 1.3686 kg/kWh, demonstrating effective utilization of both thermal and electrical energy supplied to the drying system.

Table 1. Main simulation parameters

| Parameters | Description | Value |
|---|---|---|
| Dryer volume | Effective dryer air volume | 600 [m³] |
| Drying area | Effective pulp drying area | 2700 [m²] |
| Wall area | Heat transfer area to surroundings | 900 [m²] |
| Initial pulp moisture | Initial moisture content | 0.32 [kg/kg] |
| Inlet pulp moisture | Nominal inlet moisture | 0.60 [kg/kg] |
| Initial drying-air temperature | Dryer operating temperature | 392 [K] |
| Initial pulp temperature | Web temperature | 350 [K] |
| Initial chamber pressure | Internal pressure | 101.9 [kPa] |
| Nominal airflow rate | Drying airflow | 75 [m³/s] |
| Fan inertia | Rotational inertia | 85 [kg·m²] |
| Latent heat of vaporization | Water evaporation energy | 2.257 [MJ/kg] |
| Pulp feed rate | Throughput | 8 [kg/s] |

Table 2. Closed-loop performance indicators of the proposed airborne pulp dryer control system.

| Performance index | Description | Value |
|---|---|---|
| $IAE_X$ | Integral absolute error of outlet pulp moisture tracking | 12.5627 |
| $IAE_{Ta}$ | Integral absolute error of drying-air temperature tracking | 2403.4794 |
| $IAE_{Dpsy}$ | Integral absolute error of psychrometric driving-force regulation | 9.3615 |
| $IAE_P$ | Integral absolute error of chamber-pressure regulation | 5462.4177 |
| $IAE_q$ | Integral absolute error of airflow regulation | 361.7557 |
| $E_{specific}$ | Specific energy consumption | 2.60 [MJ/kg water] |
| $SMER_{avg}$ | Average specific moisture evaporation ratio | 1.3859 [kg/kWh] |

Table 3. Final steady-state operating conditions

| Variable | Description | Final value |
|---|---|---|
| $X_p$ | Outlet pulp moisture content | 0.1201 [kg/kg dry pulp] |
| $T_a$ | Drying-air temperature | 409.83 [K] |
| $D_{psy}$ | Psychrometric driving force | 0.2039 [kg/kg] |
| $\eta_{tot}$ | Overall dryer efficiency | 0.8581 |
| $\eta_{th}$ | Thermal efficiency | 0.8697 |
| $\eta_f$ | Fan efficiency | 0.1895 |
| $SMER$ | Specific moisture evaporation ratio | 1.3687 [kg/kWh] |

Fig. 3.1 presents the principal controlled variables of the airborne pulp dryer. Fig. 3.1(a) shows the outlet pulp moisture content together with its reference trajectory. Following the startup transient, the moisture content rapidly converges toward the desired operating value and remains close to its reference despite the presence of multiple process disturbances. Fig. 3.1(b) illustrates the fan rotational speed, which exhibits a rapid initial adjustment followed by smooth convergence toward its operating point, indicating stable electromechanical behavior of the airflow generation subsystem. Fig. 3.1(c) presents the drying-air flowrate, showing that the cascade airflow controller successfully maintains the required circulation rate throughout the drying process. Fig. 3.1(d) illustrates the psychrometric driving force, which remains within the desired operating region and follows its adaptive reference, thereby preserving the thermodynamic potential required for moisture evaporation. Fig. 3.1(e) presents the drying-air temperature response. The thermal controller effectively compensates process disturbances while maintaining stable operation and negligible steady-state offset. Finally, Fig. 3.1(f) shows the chamber-pressure response, demonstrating that the pressure-control loop successfully regulates the internal pressure conditions required for stable airflow distribution and drying operation.

The internal thermo-fluid behavior of the airborne dryer is presented in Fig. 3.2. Fig. 3.2(a) illustrates the evolution of the air and pulp temperatures. The results reveal the expected thermal interaction between the hot drying medium and the moist pulp web, with both temperatures evolving toward a stable operating condition. Fig. 3.2(b) presents the heat-transfer and mass-transfer coefficients. These coefficients remain bounded and vary according to airflow conditions, confirming the strong coupling between aerodynamic operation and transport phenomena. Fig. 3.2(c) shows the air-density dynamics, which decrease slightly as the drying-air temperature increases, consistent with ideal-gas thermodynamics. Fig. 3.2(d) presents the dry-air inlet and outlet mass-flow rates, demonstrating stable circulation and conservation of the air inventory within the drying chamber. Fig. 3.2(e) illustrates the thermal power terms associated with convective heat transfer, moisture evaporation, and thermal losses. The results indicate that evaporation constitutes the dominant thermal demand of the drying process. Finally, Fig. 3.2(f) shows the duct pressure losses. The rapid decay of pressure loss after startup indicates that the airflow network reaches a stable aerodynamic regime during normal operation.

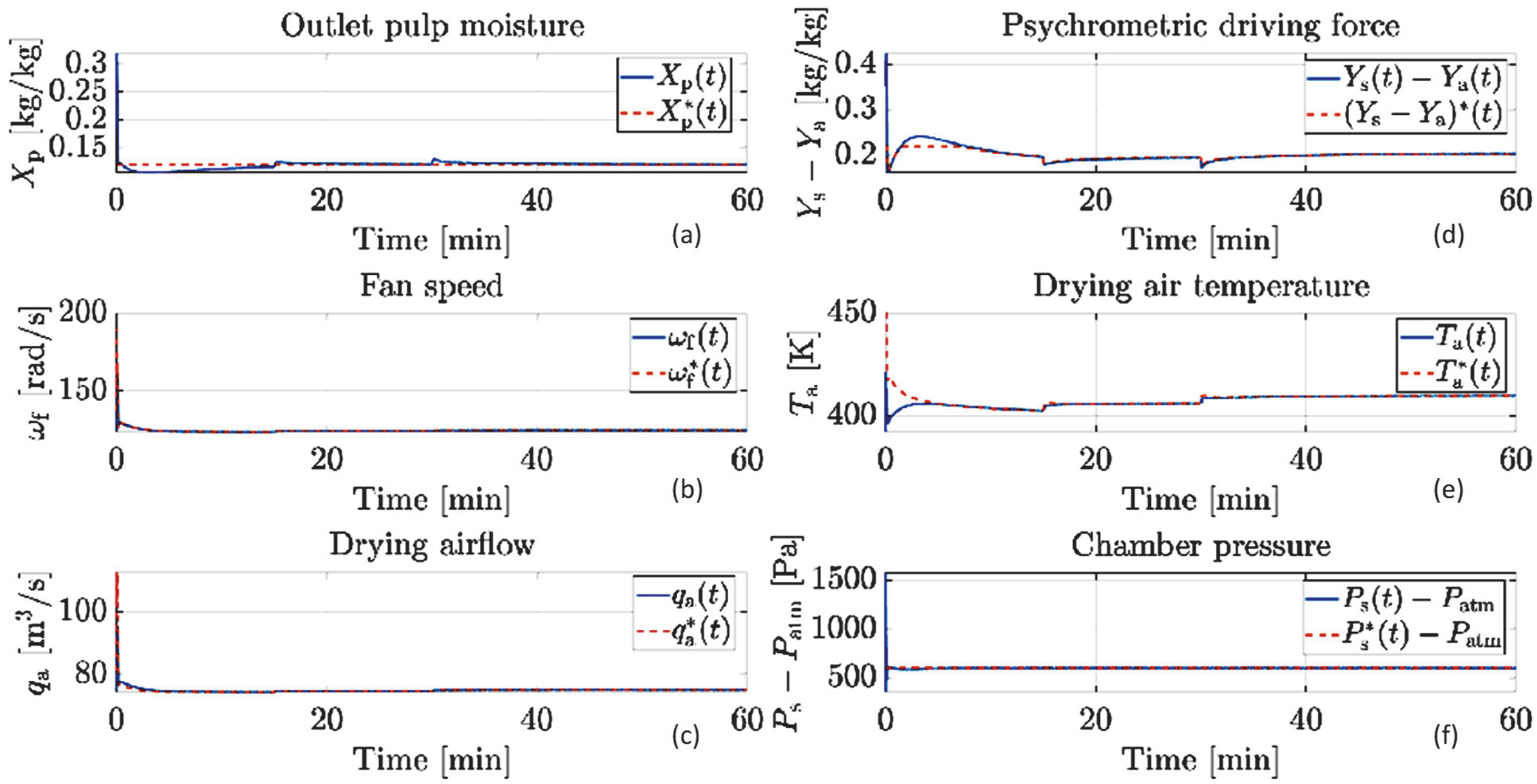

Fig. 3.1. Closed-loop dynamic responses of the airborne pulp dryer under process disturbances, showing the regulation of (a) outlet pulp moisture content, (b) psychrometric driving force, (c) fan rotational speed, (d) drying-air temperature, (e) drying-air flowrate, and (f) chamber pressure.

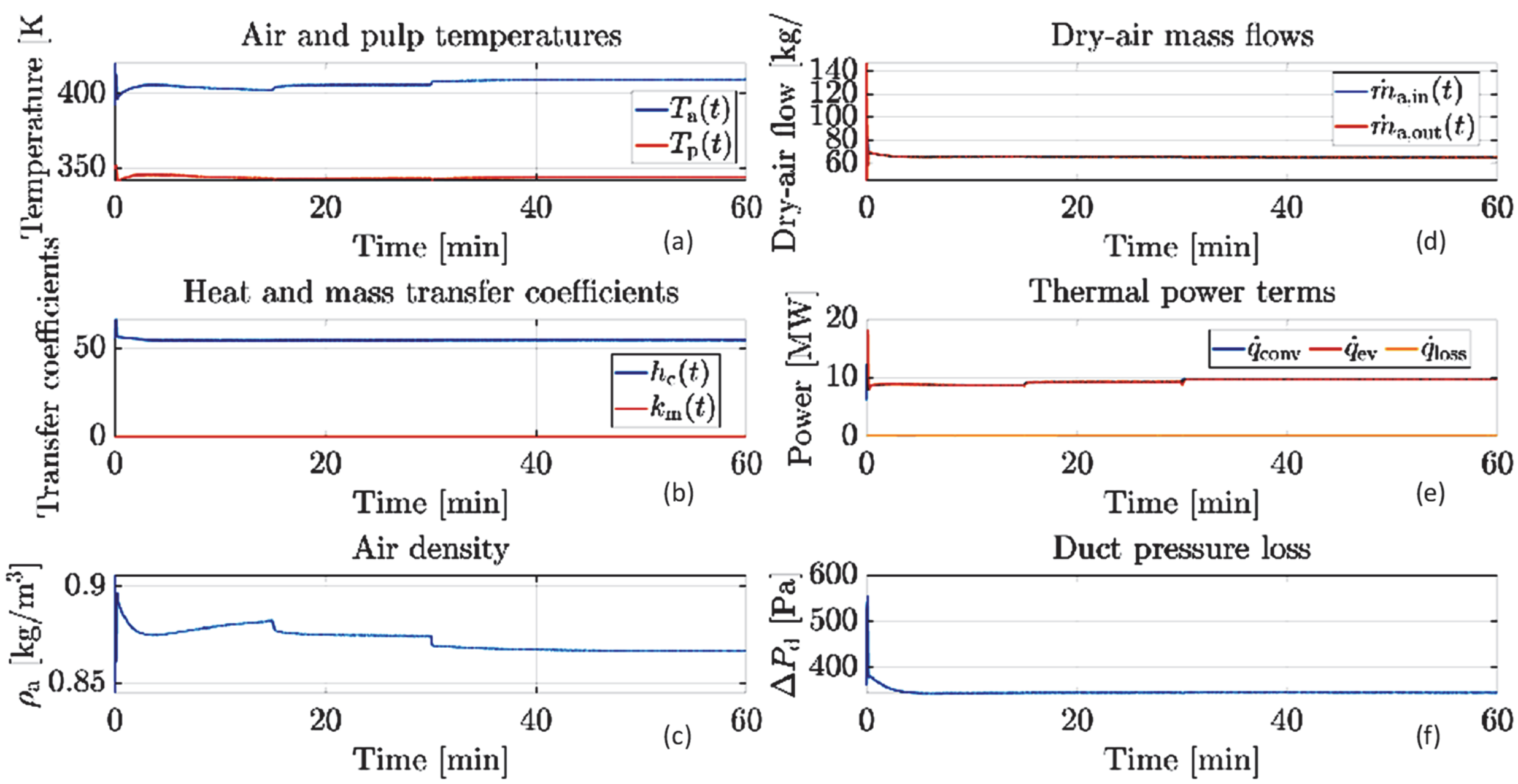

Fig. 3.2. Dynamic evolution of the internal thermo-fluid variables of the airborne pulp dryer, showing (a) air and pulp temperatures, (b) dry-air mass flows, (c) heat- and mass-transfer coefficients, (d) thermal power terms, (e) air density, and (f) duct pressure loss.

Fig. 3.3 presents the manipulated variables generated by the DSM-based hierarchical control system. Fig. 3.3(a) illustrates the thermal duty supplied by the heating subsystem. The thermal power increases as the controller compensates moisture-load changes and maintains the desired drying conditions. Fig. 3.3(b) presents the fan motor torque. A large initial control action is observed during startup, followed by a stable operating region with limited actuator effort. This behavior confirms the effectiveness of the inner fan-speed control loop. Fig. 3.3(c) illustrates the exhaust-damper area used for pressure regulation. The smooth variations of the damper opening demonstrate that the pressure-control loop maintains stable operation without excessive actuator movement.

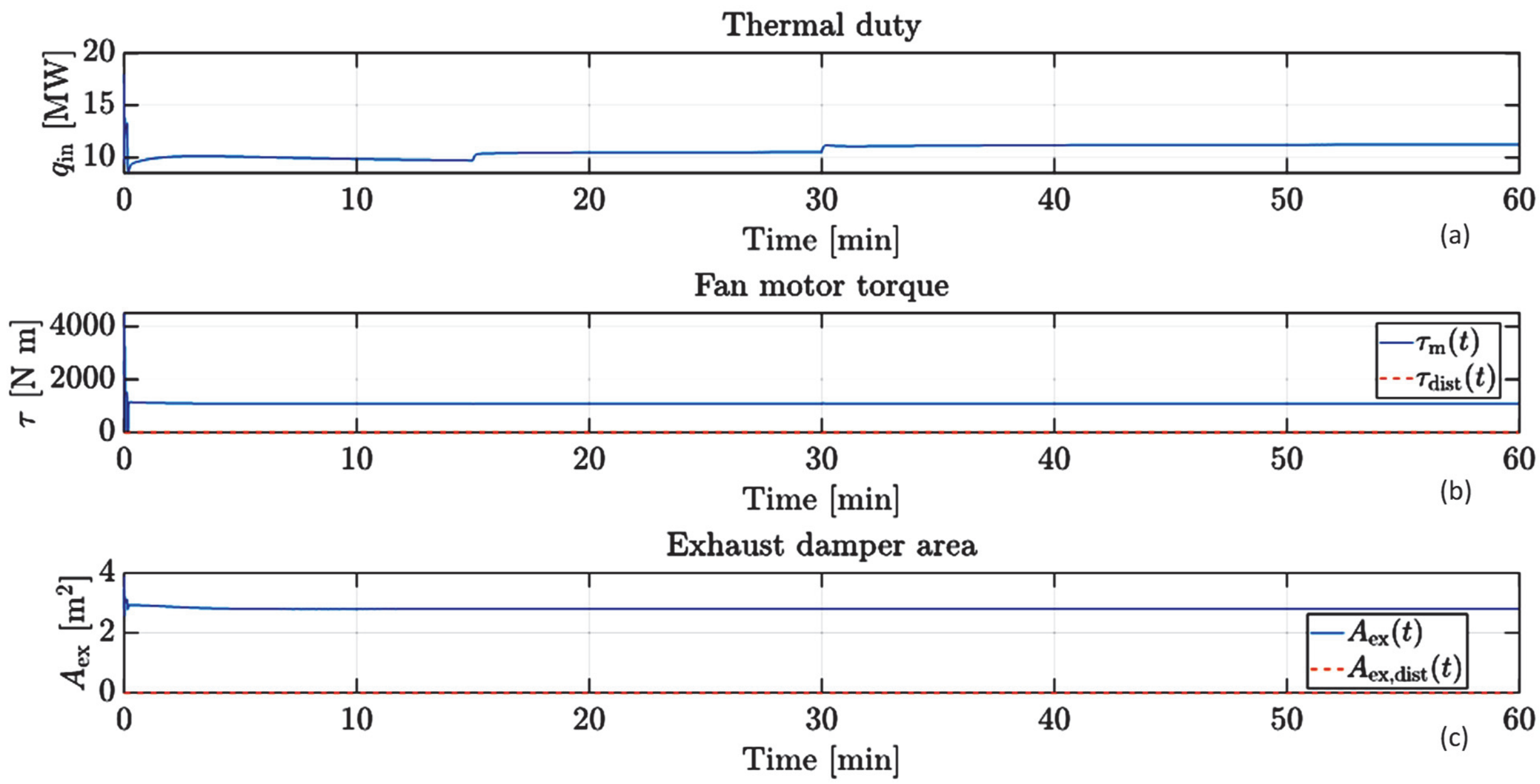

Fig. 3.3. Dynamic evolution of the manipulated variables of the airborne pulp dryer, showing (a) thermal energy input, (b) fan motor torque, and (c) exhaust damper area.

The energetic performance of the airborne dryer is summarized in Fig. 3.4. Fig. 3.4(a) presents the overall dryer efficiency. Following the initial startup transient, the efficiency stabilizes near 0.8580, indicating that the combined thermal and mechanical energy supplied to the process is effectively converted into useful moisture evaporation. Fig. 3.4(b) illustrates the thermal efficiency. The response converges toward approximately 0.8696, demonstrating that most of the supplied thermal energy contributes directly to evaporation rather than being dissipated through thermal losses. Fig. 3.4(c) presents the fan efficiency, which stabilizes near 0.1894. Although lower than the thermal efficiency, this value is consistent with industrial drying systems characterized by significant aerodynamic resistance and high airflow requirements. Fig. 3.4(d) presents the specific moisture evaporation ratio. The SMER converges toward 1.3686 kg/kWh, while the average value throughout the simulation reaches 1.3859 kg/kWh, confirming favorable energetic performance and efficient utilization of the supplied energy.

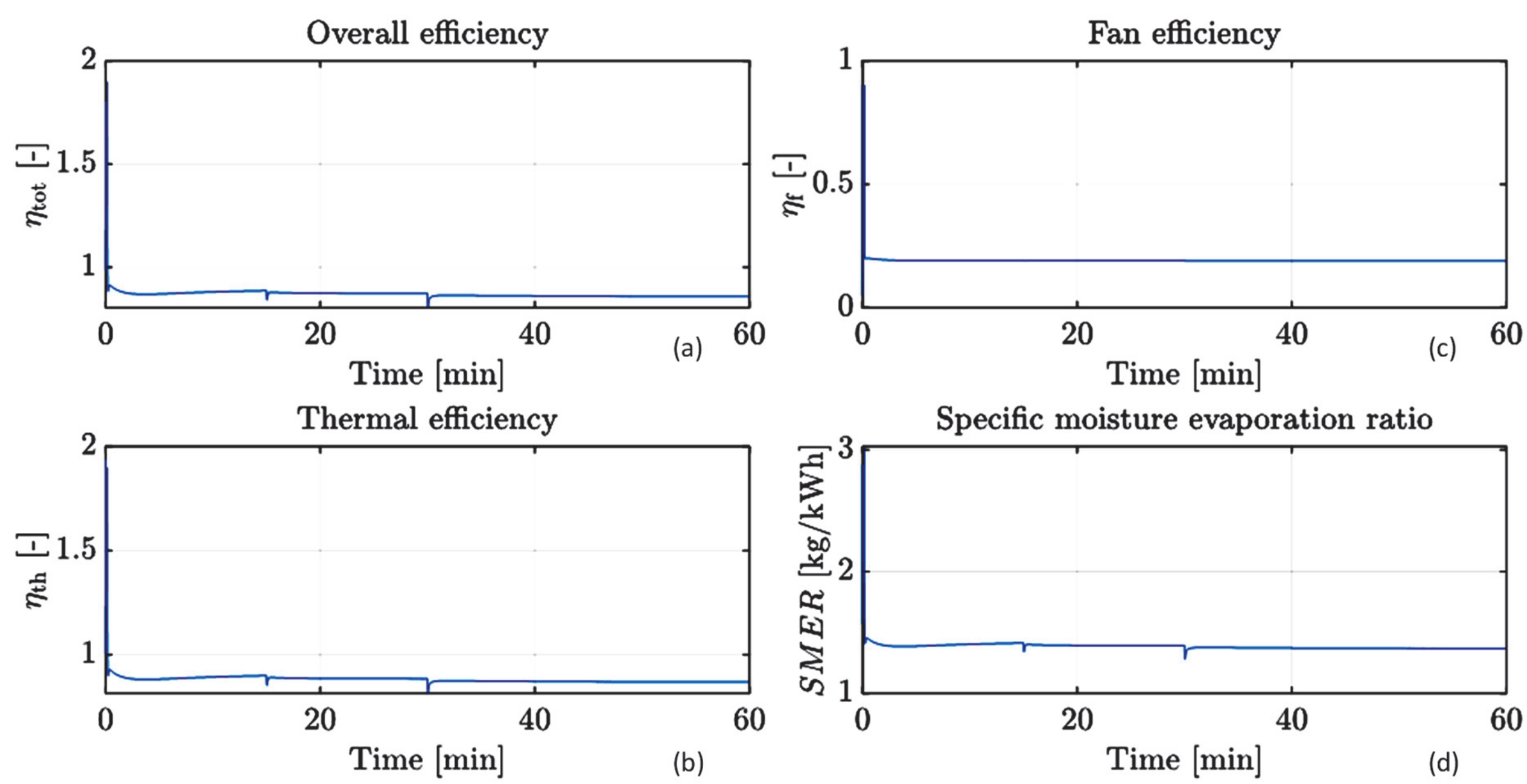

Fig. 3.4. Energetic performance assessment of the airborne pulp dryer, showing the temporal evolution of (a) overall efficiency, (b) fan efficiency, (c) thermal efficiency, and (d) specific moisture evaporation ratio (SMER).

Fig. 3.5 presents the industrial disturbances imposed during the simulation study. Figure 3.5(a) shows the inlet pulp moisture disturbance, representing variations in upstream dewatering performance. Fig. 3.5(b) illustrates the ambient-temperature disturbance, which directly affects thermal losses and the overall energy balance of the dryer. Fig. 3.5(c) presents the industrial vibration disturbance associated with fan imbalance, bearing vibration, structural oscillations, duct-flow pulsations, or drive torque ripple. Fig. 3.5(d) shows the throughput disturbance, which modifies the pulp feed rate and consequently alters residence time and evaporation demand. Fig. 3.5(e) illustrates the mechanical vibration excitation signal applied to the fan-actuator subsystem. Finally, Fig. 3.5(f) presents the resulting actuator perturbations introduced into the torque and damper channels. Despite the simultaneous presence of thermal, compositional, hydraulic, and mechanical disturbances, the proposed control architecture maintains stable operation and preserves product-quality requirements.

The multidimensional energetic behavior of the airborne dryer is further evaluated through Figs. 4.1–4.4. Fig. 4.1 presents the overall-efficiency surface as a function of drying-air temperature and airflow rate. The surface reveals a nonlinear relationship between operating conditions and energetic performance, highlighting a well-defined operating region where efficiency is maximized. Fig. 4.2 presents the thermal-efficiency surface. The results indicate that thermal utilization improves as a larger fraction of the supplied heat contributes to moisture evaporation. Fig. 4.3 shows the specific moisture evaporation ratio surface. The obtained surface identifies the operating region that maximizes water removal per unit energy supplied and confirms that maximum evaporation capacity does not necessarily correspond to maximum energetic efficiency. Finally, Fig. 4.4 presents the overall-efficiency contour map generated by the supervisory optimization layer. The contour representation clearly identifies the optimal operating region and demonstrates the importance of simultaneously regulating airflow and temperature rather than independently tuning individual variables.

Overall, the simulation results demonstrate that the proposed nonlinear airborne pulp dryer model successfully captures the coupled interactions among moisture transport, heat transfer, psychrometric behavior, airflow generation, pressure regulation, fan rotational dynamics, and energetic performance. The DSM-based hierarchical control architecture provides accurate reference tracking, robust disturbance rejection, stable actuator behavior, and high energetic efficiency. The final operating conditions of $X_p$ = 0.1201 kg/kg dry pulp, $\eta_{tot}$ = 0.8580, $\eta_{th}$ = 0.8696, and *SMER* = 1.3686 kg/kWh confirm the suitability of the proposed framework for advanced simulation studies, digital-twin development, supervisory optimization, and future industrial implementation in airborne pulp drying systems.

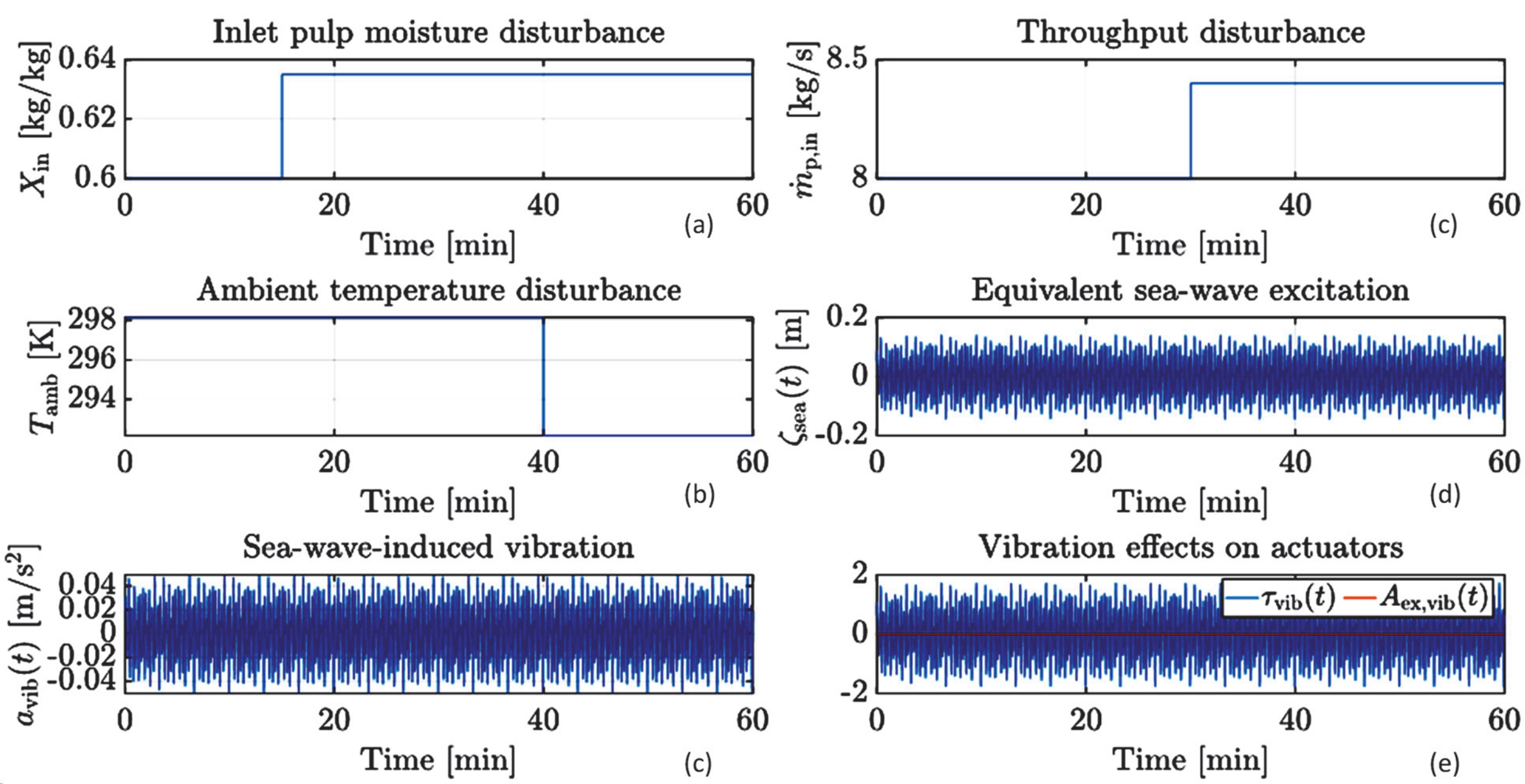

Fig. 3.5. External disturbances considered in the simulation study: (a) inlet pulp moisture disturbance, (b) ambient temperature disturbance, (c) throughput disturbance, (d) equivalent sea-wave excitation, and (e) vibration effects on the actuators.

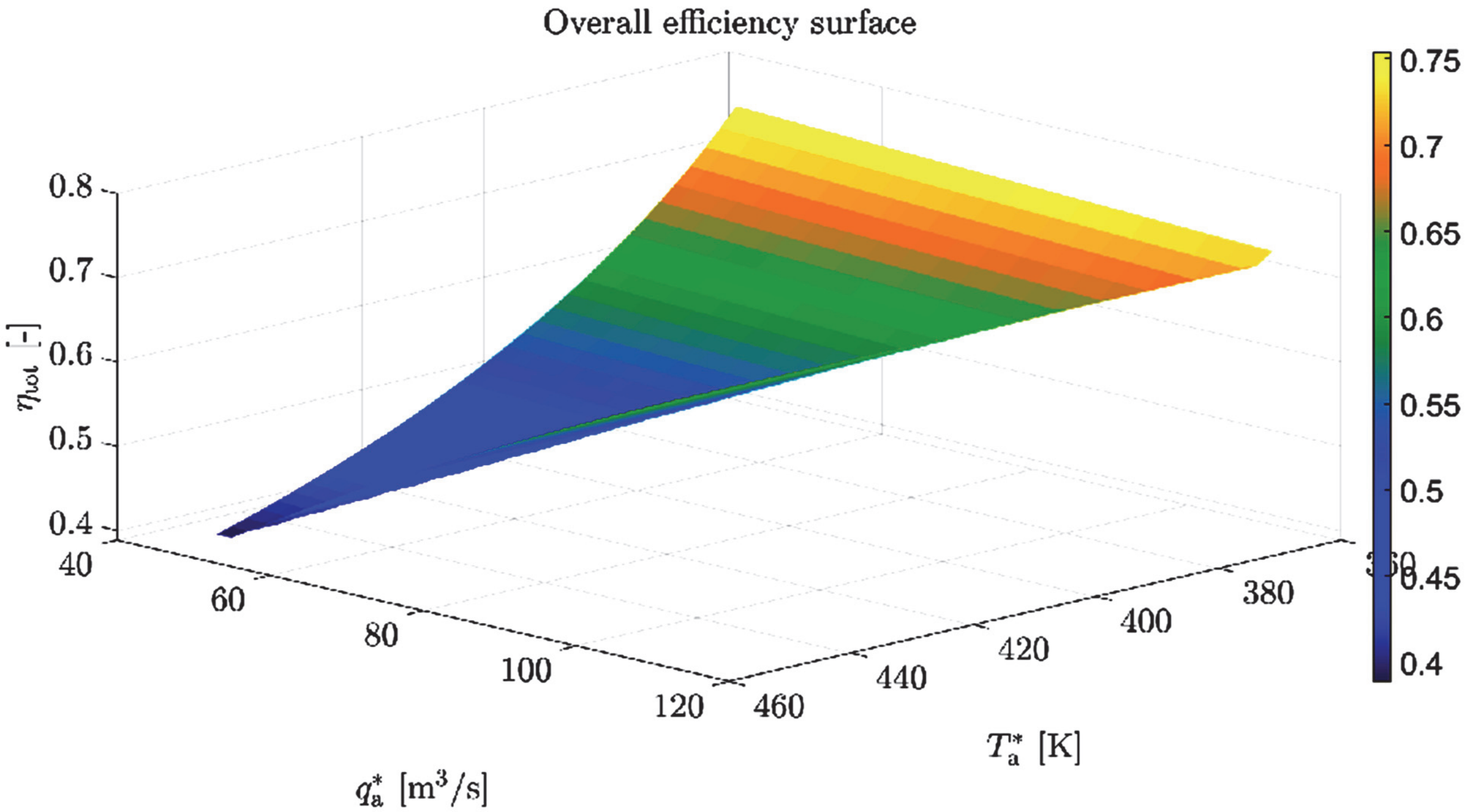

Fig. 4.1. Overall efficiency surface of the airborne pulp dryer as a function of the drying-air temperature reference $T_a^*$ and airflow reference $q_a^*$. The surface identifies the nonlinear relationship between operating conditions and energetic performance, highlighting the optimal region that maximizes the conversion of thermal and mechanical energy into useful moisture evaporation.

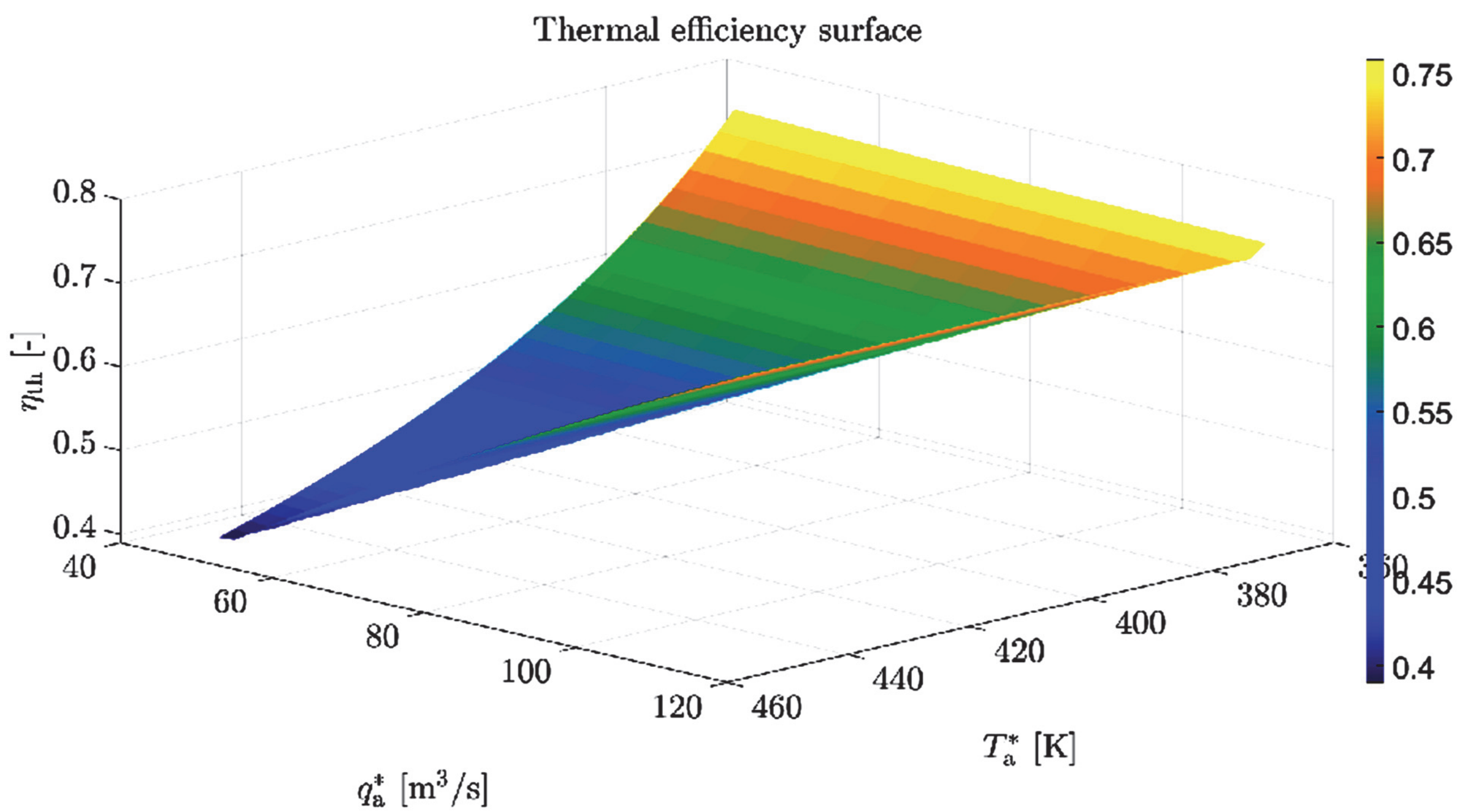

Fig. 4.2. Thermal efficiency surface of the airborne pulp dryer as a function of the drying-air temperature reference $T_a^*$ and airflow reference $q_a^*$. The surface illustrates the effectiveness of converting supplied thermal energy into useful moisture evaporation and identifies the operating conditions that maximize thermal energy utilization.

## 7.0. CONCLUSIONS

A comprehensive nonlinear dynamic modeling and control framework for an industrial airborne pulp dryer was developed and analyzed. The proposed formulation integrates the fundamental physical mechanisms governing the drying process, including dry-air transport, water-vapor accumulation, pulp moisture dynamics, heat-transfer phenomena, psychrometric behavior, airflow-pressure interactions, and centrifugal fan electromechanical dynamics. Unlike conventional steady-state or empirical approaches frequently employed in industrial drying applications, the developed model provides a unified first-principles representation capable of describing the coupled thermo-fluid, compositional, hydraulic, and mechanical interactions occurring inside modern multi-deck airborne pulp dryers. The resulting mathematical framework constitutes a physically consistent and control-oriented platform suitable for dynamic simulation, optimization, digital-twin development, and advanced control applications in pulp and paper manufacturing systems.

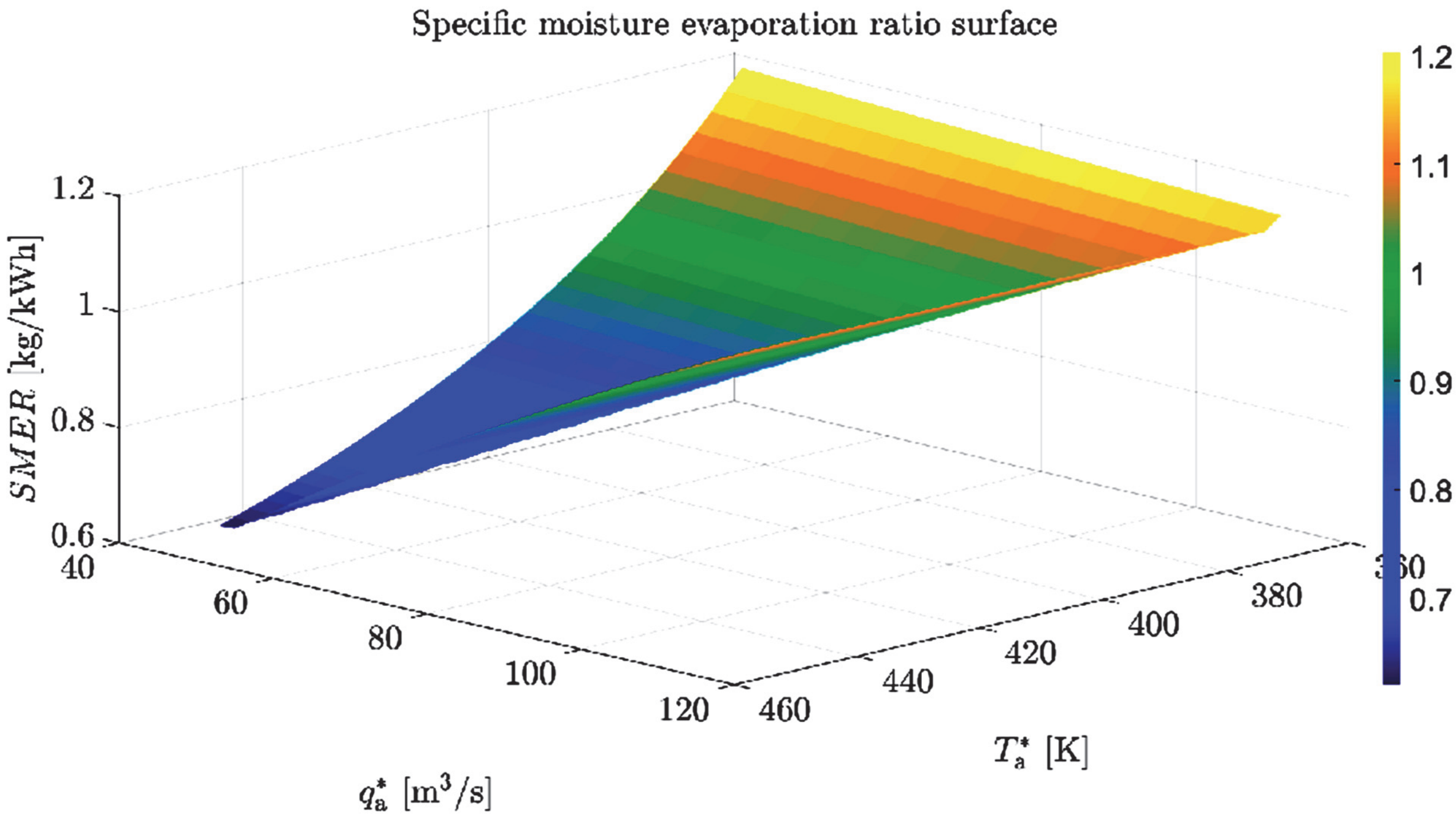

Fig. 4.3. SMER surface of the airborne pulp dryer as a function of the drying-air temperature reference $T_a^*$ and airflow reference $q_a^*$. The surface identifies the operating conditions that maximize moisture removal per unit of supplied energy, thereby providing a direct measure of the energetic performance of the drying process.

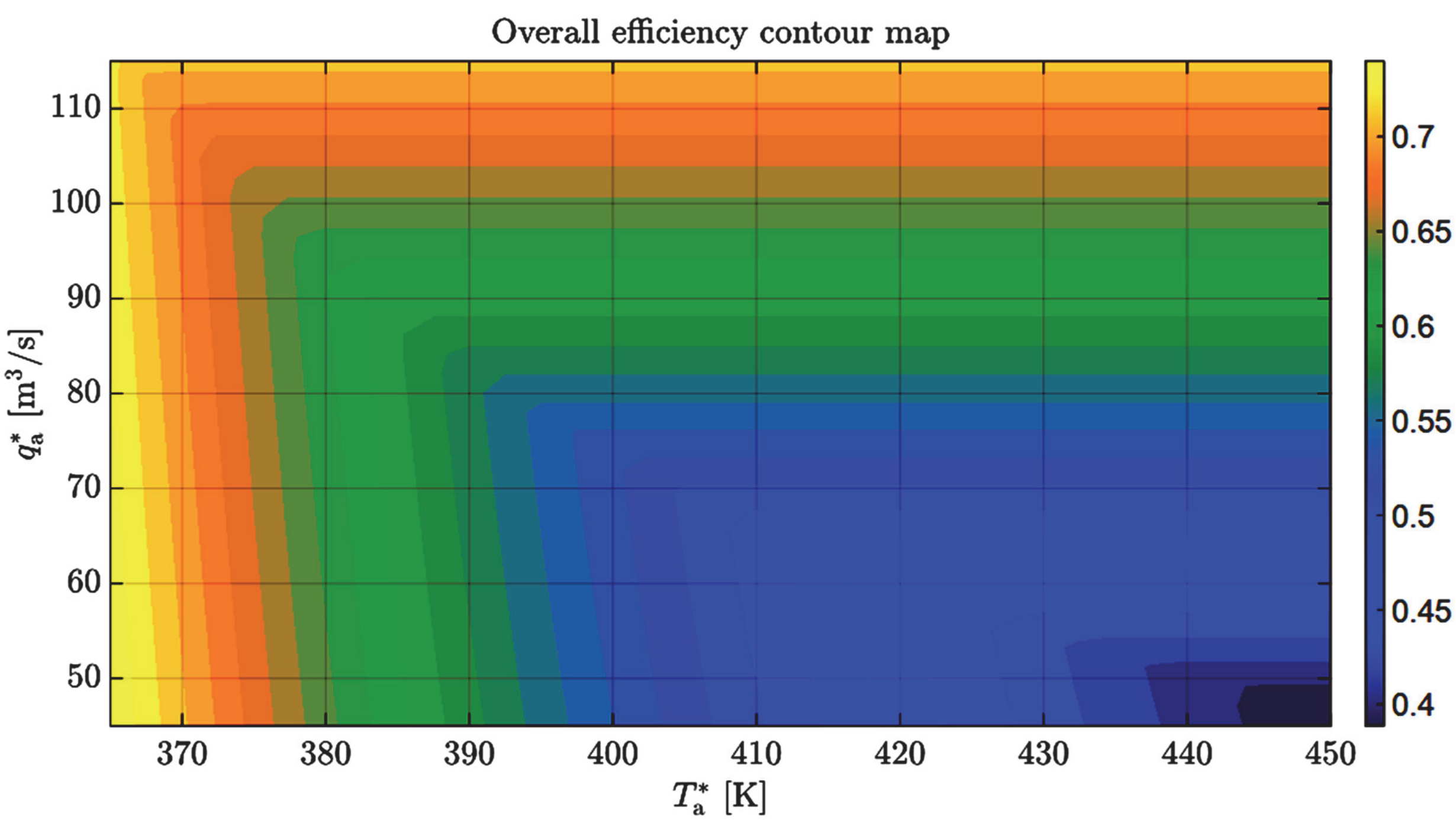

Fig. 4.4. Overall efficiency contour map of the airborne pulp dryer as a function of the drying-air temperature reference $T_a^*$ and airflow reference $q_a^*$. The contour representation identifies the optimal operating region that maximizes energetic performance and provides a practical visualization for supervisory optimization and operating-point selection.

The derived governing equations revealed the strong nonlinear coupling existing among evaporation kinetics, air humidity, pressure distribution, airflow generation, thermal transport, and moisture migration within the pulp sheet. In particular, the psychrometric driving force was shown to play a fundamental role in determining the evaporation capacity of the dryer, while the airflow generation subsystem introduced significant electromechanical interactions through the fan rotational dynamics and pressure-flow characteristics. The developed model successfully captures these interactions and provides a mechanistic explanation of the transient behavior commonly observed in industrial airborne drying systems.

An efficiency analysis framework was additionally developed to quantify the energetic performance of the drying process. The proposed formulation incorporates overall efficiency, thermal efficiency, fan efficiency, and SMER, thereby enabling a multidimensional evaluation of energy utilization within the dryer. Unlike conventional analyses that focus exclusively on thermal efficiency, the proposed approach explicitly accounts for both thermal and mechanical energy consumption, providing a more realistic assessment of industrial drying performance. The resulting efficiency expressions demonstrated that maximizing moisture removal alone does not necessarily lead to optimal energetic performance and that coordinated regulation of airflow, temperature, pressure, and evaporation dynamics is required to achieve energy-efficient operation.

A hierarchical control architecture was subsequently developed based on the DSM. The proposed strategy combines cascade moisture–temperature control, cascade airflow–fan speed regulation, an independent pressure-control loop, and an efficiency-based supervisory optimization layer. This architecture was specifically designed to address the multivariable nature of the airborne drying process while simultaneously satisfying product-quality requirements and energetic performance objectives. The supervisory layer continuously evaluates the efficiency indicators and adjusts the drying-air temperature and airflow references in order to maximize the overall dryer performance under varying operating conditions.

The simulation results demonstrated the effectiveness of the proposed control strategy. The outlet pulp moisture converged to 0.1201 kg/kg dry pulp, confirming accurate regulation of the primary quality variable. Similarly, the drying-air temperature reached 409.83 K and the psychrometric driving force stabilized at 0.2039 kg/kg, providing favorable thermodynamic conditions for sustained evaporation. The obtained tracking performances, characterized by $IAE_{\mathrm{X}}$ = 12.5624, $IAE_{\mathrm{Dpsy}}$ = 9.3609, and $IAE_{\mathrm{q}}$ = 361.3310, demonstrate satisfactory closed-loop regulation despite the presence of significant thermal, compositional, hydraulic, and mechanical disturbances.

From an energetic perspective, the proposed dryer-control system achieved highly favorable performance. The overall efficiency converged to 85.80%, while the thermal efficiency reached 86.96%. Furthermore, the average SMER attained 1.3859 kg/kWh, with a final steady-state value of 1.3686 kg/kWh. These results indicate that the majority of the supplied thermal energy was effectively converted into useful moisture evaporation, while fan energy consumption remained within acceptable industrial limits. The specific energy consumption of 2.60 MJ/kg of evaporated water further confirms the energetic consistency of the proposed drying strategy.

The multidimensional efficiency surfaces provided additional insight into the operational characteristics of the dryer. The generated efficiency maps revealed the existence of well-defined optimal operating regions governed by the interaction between airflow rate and drying-air temperature. These results demonstrate that the energetic behavior of airborne pulp dryers is inherently nonlinear and that optimal operation cannot be achieved through independent tuning of individual process variables. Instead, coordinated multivariable optimization is required to maximize moisture removal while minimizing thermal and electrical energy consumption.

The disturbance analysis further demonstrated the robustness of the proposed control architecture. Variations in inlet pulp moisture, ambient temperature, production throughput, industrial vibration disturbances, and actuator perturbations were successfully rejected without compromising process stability or product quality. The controlled variables remained bounded throughout the simulation horizon, and no significant oscillatory behavior was observed, confirming the suitability of the DSM-based cascade and non-cascade control structures for industrial airborne drying applications.

One of the most significant contributions of this work lies in the development of what is believed to be one of the first comprehensive first-principles dynamic frameworks for airborne pulp dryers that simultaneously incorporates mass balances, energy balances, psychrometric relationships, airflow-pressure dynamics, moisture transport phenomena, and fan electromechanical behavior within a unified modeling environment. While airborne drying technologies are extensively employed throughout the pulp and paper industry, publicly available dynamic models of this complexity remain extremely limited in the open scientific literature. Consequently, the proposed framework provides a valuable foundation for future research and industrial implementation.

Future work may focus on extending the proposed framework toward distributed-parameter modeling, computational fluid dynamics integration, moisture-diffusion modeling within the pulp fiber network, advanced airflow recirculation strategies, and model predictive control implementations. Furthermore, experimental validation using industrial airborne dryers would provide additional insight into parameter identification and model calibration. The proposed framework may also serve as the basis for digital-twin platforms, real-time optimization systems, energy-management strategies, and next-generation intelligent drying systems for sustainable pulp and paper manufacturing.

Overall, the developed nonlinear dynamic model, efficiency formulation, and hierarchical control architecture constitute a rigorous and comprehensive platform for the analysis, simulation, optimization, and control of industrial airborne pulp dryers. The obtained results demonstrate that the proposed methodology is capable of simultaneously achieving stable moisture regulation, efficient energy utilization, robust disturbance rejection, and favorable dynamic performance, thereby providing a technically sound foundation for future developments in advanced pulp drying technologies.